\documentclass[article,prb,twocolumn,floatfix,showpacs]{revtex4-1}
\usepackage{epsfig,subfigure,amssymb,amscd,amsmath,graphicx,amsfonts}
\usepackage{times,longtable}

\newcommand{\non}{\nonumber}

\newcommand{\bea}{\begin{eqnarray}}
\newcommand{\eea}{\end{eqnarray}}
\newcommand{\be}{\begin{equation}}
\newcommand{\ee}{\end{equation}}
\newcommand{\ba}{\begin{align}}
\newcommand{\ea}{\end{align}}

\newcommand{\ZZ}{\mathbb{Z}}

\renewcommand{\vr}{{\boldsymbol{r}}}

\newcommand{\bphi}{{\boldsymbol{\phi}}}

\usepackage[usenames]{color}

\newcommand{\rf}[1]{(\ref{#1})}

\begin{document}

\title{Kitaev spin models from topological nanowire networks}

\author{G. Kells$^{1,2}$, V. Lahtinen$^{3}$, J Vala$^{1,4}$ }

\affiliation{$^{1}$ Dept. of  Mathematical  Physics,  National University of Ireland, Maynooth,~Ireland. 
\\$^{2}$ Dahlem Center for Complex Quantum Systems and Fachbereich Physik, Freie Universit\"{a}t Berlin, Arnimallee 14, 14195 Berlin, Germany.
\\$^{3}$ Institute for Theoretical Physics, University of Amsterdam, Science Park 904, 1090 GL Amsterdam, The Netherlands.
\\$^{4}$ Dublin Institute for Advanced  Studies, School of Theoretical Physics, 10 Burlington Rd, Dublin, Ireland.}

\begin{abstract}
We show that networks of superconducting topological nanowires can realize the physics of exactly solvable Kitaev spin models on trivalent lattices. This connection arises from the low-energy theory of both systems being described by a tight-binding model of Majorana modes. In Kitaev spin models the Majorana description provides a convenient representation to solve the model, whereas in an array of Josephson junctions of topological nanowires it arises from localized physical Majorana modes tunnelling between the wire ends. We explicitly show that an array of junctions of three wires -- a setup relevant to topological quantum computing with nanowires -- can realize the Yao-Kivelson model, a variant of Kitaev spin models on a decorated honeycomb lattice. Employing properties of the latter, we show that the network can be constructed to give rise to two dimensional collective topological states characterized by Chern numbers $\nu = 0, \pm 1$ and $\pm 2$, and that defects in the array can be associated with vortex-like quasi-particle excitations. In addition we show that the collective states are stable in the presence of disorder and superconducting phase fluctuations. When the network is operated as a quantum information processor, the connection to Kitaev spin models implies that decoherence mechanisms can in general be understood in terms of proliferation of the vortex-like quasi-particles. 
\end{abstract}


\date{\today} \maketitle

\section{Introduction} 
\label{sect:Intro} 
The prospect of quantum computation has spurred research into physical systems that could offer sufficient stability and control to carry out qubit manipulations in a robust manner. Topological quantum computation -- an initially exotic idea of using topological properties of materials -- has recently emerged as a serious contender. The breakthrough was the discovery that topological insulators in proximity to a standard s-wave superconductor provided a relatively simple route to realize the central element of such proposals:\cite{Fu2008} localized Majorana quasi-particles with non-Abelian statistics. It was soon realised that the essential physics could also be achieved in a simpler setting, namely with conventional semiconductors with spin-orbit coupling.\cite{Sau2010,Alicea2010}  From the perspective of scalable quantum computation a key element was the subsequent discovery that a 1D topological p-wave superconductor, originally considered as a toy model that supports Majorana bound-states,\cite{Kitaev2001,Motrunich2001} could be effectively realised using the spin-orbit coupled semiconductor nanowires.\cite{Lutchyn2010,Oreg2010} These studies were followed by proposals to braid the Majorana end states, which demonstrated that topological nanowire networks could in principle support the essential components of topological quantum computation.\cite{Alicea2011}  Recently  experiments on the nanowires have been carried out with the results supporting the existence the Majorana modes.\cite{Mourik2012,Das2012,Churchill2013} While loophole-free evidence still awaits, \cite{Liu2012,Bagrets2012,Pikulin2012,Kells2012,Lee2012} it seems plausible that Majorana modes will become a reality. 

An essential component of topological nanowire based schemes of topological quantum computation is the T-junction -- a Josephson junction where three topological nanowires come into proximity -- which can be used to braid and manipulate the Majorana end states.\cite{Alicea2011,Sau2011,vanHeck2012,Halperin2012,Fulga2013}  A scalable architecture for topological quantum computer would consist of many of these junctions brought together in a regular array.\cite{Hyart2013} One may then wonder whether the microscopics of the system when confined to a finite volume may affect the nature of the array. Indeed, the Majorana modes are localized exponentially, which means that they can tunnel between the wire ends. For sparse arrays this leads to exponential degeneracy lifting that gives a source of decoherence. For dense arrays, however, something more dramatic could happen: the Majorana modes could hybridize and form another collective topological state, very much like what can happen in Majorana mode binding vortex crystals\cite{Lahtinen2012}. This would require going through a phase transition, resulting in the significant degredation of the encoded information. Thus it is important to understand under what circumstances such collective states can form in topological wire arrays.

This question was addressed in a different setting\cite{Xu2010,Terhal2012,Nussinov2012}, where it is shown that when wires are deposited on top of an array of superconductors, the Majoranas become interacting and various collective states can emerge. In this work our focus is closer to the proposals for braiding the Majorana end states\cite{Alicea2011,Sau2011} where nanowires are placed in proximity to a common superconductor.  We show that the low-energy theory of a static array is described by an effective tight-binding model of free Majorana fermions subject to two distinct tunnellings: intra-wire tunnelling along the nanowires and a fractional Josephson tunnelling between the nanowires\cite{Kwon2004}. Our main result is to show that the low-energy theories of various wire arrays realize parts of the phase diagrams of the class of spin 1/2 lattice models which are collectively known as Kitaev spin models. The original model was defined  on a honeycomb lattice\cite{Kitaev2006}, but they are readily generalized also to other trivalent lattices.\cite{Yao2007,Yang2007,Tikhonov2010, Kells2011,Whitsitt2012}

The connection between the wire arrays and the spin models is based on the simple observation that the latter also admit description in terms of free Majorana fermions.\cite{Kitaev2006} Finding then the correspondence between the wire array tunnelling amplitudes and those of the corresponding spin model enables one immediately to read off the phase diagram for the array as well as apply known results about the stability of those phases under disorder.\cite{Willans2010,Willans2011,Chua2011,Lahtinen2013} We will show that if the fractional Josephson tunnelling can be made comparable in strength to the intra-wire couplings, stable collective topological states characterized by Chern numbers $|\nu|>0$ can emerge, with the precise nature of the state depending on the array geometry.

While avoiding the formation of collective states is of interest to quantum computations with wire arrays, one could also think of the wire array as a potential quantum simulator for the full range of many-body physics known to occur in Kitaev spin models. For instance, one could study the characteristic vortex interactions\cite{Lahtinen2011} that can lead to a nucleation transition when a vortex crystal forms\cite{Lahtinen2010,Lahtinen2012}, the emergence of a disorder induced thermal metal state unique to Majorana modes \cite{Laumann2012,Lahtinen2013}, the non-Abelian statistics of the vortices \cite{Lahtinen2009, Bolukbasi2012} or  impurity effects. \cite{Willans2010, Willans2011, Dhochak2010} Thus we believe that topological nanowire arrays are not only interesting from the point of view of their potential for topological quantum computing, but that as the experiments become more sophisticated, they could also contribute more generally to the understanding of topological condensed matter.

Our paper is structured as follows. In Section \ref{sect:network} we review the elementary building block, the $p$-wave superconducting nanowire. We will derive the effective Majorana hopping model that arises when $N$ such wires are brought together to form an  $N$-junction and subsequently arranged on a regular array.  In Section \ref{sect:Kitaev} we review the solution and the general vortex sector structure of Kitaev spin lattice models. Section \ref{sect:YK} forms the main body of our work. There we first explicitly demonstrate the equivalence between the 3-junction array and the Yao-Kivelson variant of the Kitaev spin lattice models.  By numerically solving a full microscopic model for the wire array (given in Appendix \ref{sect:App_micromodel}), we demonstrate that the effective Majorana model indeed provides an accurate description of the system. We will also show that while only Chern number $\nu=\pm 1$ phases are obtainable in regular arrays, higher Chern numbers can in principle be obtained by either creating effective vortex lattices or by considering $N>3$ junction arrays (details are given in Appendix \ref{App_Higher}). Finally, in Section \ref{sect:Stability} we discuss the stability of the collective states in the wire arrays. The correspondence between the decoherence mechnisms in the arrays as topological quantum computers and the quasiparticle dynamics in the collective states is further discussed in Appendix \ref{sect:FVS}.

\section{The N-junction wire network}
\label{sect:network} 

In this section we first review the elementary building block of a wire network -- the superconducting $p$-wave wire that hosts localized Majorana end states. Then we bring $N$ such wires together to form a Josephson junction and review the collective behavior of the end states due to the fractional Josephson physics resulting from single electron tunnelling. Finally, we arrange the junctions in a periodic array and argue that the low-energy physics of the array can be described by a tight-binding model for the Majorana end states.

\subsection{The spinless $p$-wave wire}
A basic element of a wire array is a single $p$-wave paired nanowire. There are numerous proposals for realizing them in microscopically distinct systems, such as topological insulators,\cite{Fu2008} semiconductor wires,\cite{Oreg2010,Lutchyn2010} half-metals,\cite{Duckheim2011,Qi2011} cavity arrays,\cite{Bardyn2012} nano-particles,\cite{Choy2011} or magnetic molecules\cite{NadjPerge2013}.  Regardless of the implementation though, the low-energy physics can always be expressed in the form of a simple 1D $p$-wave superconducting model first explored by Kitaev \cite{Kitaev2001} and Motrunich {\em et al.} .\cite{Motrunich2001} The continuum limit the Hamiltonian of this model can be written as
\be 
\non
H=  \int \Psi^\dagger (x) H^{}_{BdG} \Psi(x) dr
\ee
with $\Psi^\dagger(r) = \left[\begin{array}{c} \psi^\dagger(r) , \psi(r)\end{array}  \right]$ and
\be \label{H1Dcont}
H^{}_{BdG} = [ \frac{p^2}{2 m} -\mu(x) +V(x)] \tau^z - \Delta(x)  p \tau^y,
\ee
where $\tau^\alpha$ are the usual Pauli matrices. The electron mass $m$, the chemical potential $\mu(x)$, the pairing term $\Delta(x) = |\Delta| e^{ i \bphi(x)}$ and the confining potential $V(x)$ will in general depend on the microscopic realization, but here we will treat them as independent parameters. The relevant derived parameters are the superconducting energy gap, Fermi momentum and the coherence length, which are given by $\Delta_E = |\Delta| k_F$, $k_F =\sqrt{2 m \mu}$ and $\xi= 1 / m |\Delta| $, respectively. 

We model a wire of length $L$ by setting the relative values of the chemical potential and the confining potential as follows
\be
\begin{array}{rclrclcl} \label{V}
  V(x) & = & 0, & \mu(x)& =&\mu, & \qquad & 0 \leq x \leq L, \nonumber \\
  V(x) & = & V_0, & \mu(x)&=&0, & \qquad & x < 0 \ \textrm{or} \ x > L. \nonumber
\end{array}
\ee
When $|\Delta|>0$ and $\mu > 0$ the wire is known to be in a topological phase with a Majorana modes exponentially localized at each end of the wire\cite{Kitaev2001}. In the limit $L\rightarrow \infty$ the Majorana modes have precisely the energy $E=0$ and the corresponding operators are explicitly given as
\bea \label{eq:Majdef}
\gamma_b(x) & = & \frac{1}{\sqrt{2 \mathcal{N}}} \left[  e^{~i\phi/2} \psi(x)^\dagger + e^{-i\phi/2} \psi(x) \right] u(x), \\
\gamma_w(x) & = & \frac{i}{\sqrt{2 \mathcal{N}}} \left[ e^{~i\phi/2} \psi(x)^\dagger - e^{-i\phi/2} \psi(x) \right] u(L-x), \nonumber
\eea
where we denote the $x=0$ ($x=L$) end of the wire as black $(b)$ (white $(w))$ and $\mathcal{N}$ is a normalization factor. The wavefunction $u(x)$ depends whether it extends into the non-topological ($x<0$ and $x>L$) or topological ($0 \leq x \leq L$) region. In these two distinct cases it is given by
\bea
u(x) &=& A e^{x/\xi_J}, \\
u(x) &=& B  e^{-x/\xi + i x \bar{k}_F  } + C e^{-x/\xi -i x \bar{k}_F}, \nonumber
\eea
respectively. Here $\bar k_F = \sqrt{k_F^2 - 1/\xi^2}$ and we defined $\xi_J=1/\sqrt{2 m V_0}$ as the decay length into the non-topological region. As we only consider situations where the Fermi wavelength $\lambda_F = 2 \pi/ k_F$ is much smaller than $\xi$, we will approximate $\bar{k}_F= k_F$. 

The precise form how the wavefunctions decay into different regions depends on the potential $V_0$ that describes the magnitude and shape of the energy barrier due to the junction. In general, it depends on the microscopics of the system realizing the $p$-wave nanowire. However, to study the collective behavior of the wire array, we adopt initially an idealistic picture where each end of the wire is terminated in a hard wall manner ($V_0 \gg \Delta_E $ and it is of the step function form \rf{V}). In this situation one can choose $B=-C=-i/2$ and $A=0$, which means the wavefunction will decay only to the topological region 
\bea
u(x) & = & 0, \qquad \qquad \qquad \quad x<0, \\
u(x) & = & \sin ( k_F x ) e^{-x/\xi} \qquad x>0. \nonumber
\eea
Under this hard wall approximation the ground state manifold of a single wire in the topological phase will contain two states that correspond to the occupation $d^\dagger d = (1+i\gamma_b\gamma_w)/2$  of the delocalized single fermion mode $d=(\gamma_b+i\gamma_w)/2$ shared by the two localized Majoranas. When the wire is of infinite length these states have zero energy and they are separated by all other states in the spectrum by the energy gap $\Delta_E$. When the wire is finite and/or the boundary conditions are more realistic (spatially smooth), the overlap between wavefunctions from two ends results in the degeneracy of the states being only exponential in the wire length. This also implies that typically one does not find states of the form (\ref{eq:Majdef}), but the form of the wave functions is slightly modified. A rigorous solution would involve solving for the low-energy spinor wavefunctions with the correct boundary conditions at both ends of the wire. However, we will employ a simpler approach by 
taking the hard wall solutions (\ref{eq:Majdef}) as ansatz states and treat the finite length and the more realistic boundary conditions as perturbations that couple them. This picture enables to view all sub-gap dynamics as Majoranas modes tunnelling between the wire ends. The effective low-energy Hamiltonian describing this is given by
\be \label{HZ}
 H' =i J' \gamma_{b}\gamma_{w} +\mbox{h.c.} 
\ee
with $J' \sim 2 \Delta_E  \sin(k_F L) e^{-L/\xi}$. The tunnelling amplitude $J'$ follows directly from the exponentially vanishing Majorana wavefunctions on the opposite hard wall terminated ends, see for example Ref. \onlinecite{Pientka2013}. Away from the hard-wall limit, i.e. when $V_0$ is finite and smooth in space, the tunnelling amplitude is only modified on the order of $\sqrt{\mu/V_0}$ .\cite{Pientka2013} Thus we assume that \rf{HZ} will provide a good approximation also for more realistic scenarios that are required to couple the end states from different wires.

\subsection{The $N$-junction of topological nanowires}

When two superconducting wires are brought into proximity, they will form a Josephson junction where a current will flow due to the tunnelling of Cooper pairs whose amplitude depends on the relative superconducting phases on each wire. When the wires are in a topological phase with Majorana modes localized at their ends, tunnelling of also single electrons is possible and one obtains a fractional Josephson junction where the tunnelling amplitude now depends on half the relative superconducting phase difference \cite{Kwon2004}. 

As above when considering the coupling between the Majoranas in the same wire, this process can be described in terms of Majorana tunnelling through a potential barrier of height $V_0$, with additional tunnelling modulation coming from the Josephson physics. This is governed by the Hamiltonian
\be \label{HJ}
H=i J  \gamma_{b/w} \gamma_{b/w} + \mbox{h.c.}, 
\ee
with $J=\Delta_E \sqrt{T} \sin(\delta\phi) $ where $T$ is the transmission coefficient at $k_F$ between different wires, and $\delta \phi  = \phi_1/2-\phi_2/2$ is half the difference of the superconducting phase in the two wires \cite{Kwon2004}. Here we adopt the convention that the wire ends meeting at a junction carry the same end label and that the superconducting phases are defined with respect to the junction. Assuming a junction of width $W$, with this region modelled as a square potential of height $V_0$, the tunnelling amplitude $J$ is given by $J=\Delta_E  \sin(\delta\phi) e^{-W/\xi_J}$.

Junctions can also be formed when more than two wires are brought into proximity. When $N$ topological nanowires form a junction, pairwise Josephson tunnelling will take place between all the wire ends and the Hamiltonian describing the junction generalizes to
\be \label{HJN}
H_N=i \sum_{n<m,m=1}^N J_{nm} (\gamma_{n,b/w}\gamma_{m,b/w})+ \mbox{h.c.}.
\ee
For $N=3$ the junction will be of the $T$-junction type, which will be important to us in the following chapter. There all the couplings $J_{nm}$ can be chosen equal given that the wire ends are equispaced in the junction. For $N>3$ this is not in general possible due to geometrical reasons that require some pairwise junctions to be wider and thus the corresponding amplitudes smaller. 

\subsection{A periodic network of $N$-junctions}

When the $N$-junctions are arranged on a two dimensional periodic array such that neighbouring junctions always alternate between black and white, the low-energy theory of the system is governed by the Hamiltonian
\be
\mathbf{H}_N = \sum_{\textrm{wires}} H' + \sum_{\textrm{junctions}} H_N + \mathcal{O}(J \times J'). \nonumber
\ee
The $\mathcal{O}(J \times J')$ terms describe exponentially weaker coupling between Majoranas end states that belong to different wires and different junctions. As second order terms in exponentially vanishing couplings $J$ and $J'$, these terms provide only small quantitative corrections which are negligible from the point of view of the general form of the phase diagram. We have verified this numerically (see Appendix \ref{sect:App_micromodel}) and will mostly neglect them from now on.

Viewing then the wire ends as the sites $i$ of a two dimensional lattice, the Hamiltonian above is then formally equivalent to the Majorana tight-binding model
\be \label{HN}
  \mathbf{H}_N = i \sum_{(i,j) \in \textrm{wires}} J'_{ij} \gamma_i \gamma_j + i \sum_{(i,j) \in \textrm{junctions}} J_{ij} \gamma_i \gamma_j.
\ee
The simplest 2D array occurs for $N=3$ when the wire ends form a decorated honeycomb lattice (sites replaced by triangles), as illustrated in Fig. \ref{fig:YK}. In the absence of the $\mathcal{O}(J \times J')$ couplings the Majorana tunnelling will be purely of nearest neighbour type with the first and second term in \rf{HN} describing Majorana tunnelling between and within the triangles, respectively. This array will be central to our discussion below. Other arrays with $N>3$ junctions are illustrated in Fig. \ref{fig:arrays}. The higher junction valency implies that some longer range tunnelling always present in the corresponding tight-binding model. These can lead to more complex phase diagrams as we will study later.

The parameters of the hopping model are given as:
\bea \label{ZJ}
  J'_{ij} & = & \Delta_E  \sin(k_F L_{ij}) e^{-L_{ij}/\xi}, \\
  J_{ij} & = & \Delta_E \sin(\delta\phi_{ij}) e^{-W_{ij}/\xi_J},
\eea
In an ideal situation all these parameters are freely tunable locally and independent of each other. This would be the case if it were possible to couple each wire to an independent $s$-wave superconductor (to tune $\Delta$ that controls $\xi$) and to an independent voltage bias gate (to tune $\mu$ controls $k_F$). While this may be possible (see for example Ref. \onlinecite{vanHeck2012}) as the experiments become more sophisticated, here we imagine a simpler setting. We assume that all the wires are deposited on top of the same superconductor, which implies that both the induced superconducting gap $\Delta_E$ and the Fermi momentum $k_F$ will be assumed to be equal in all wires. Furthermore, we assume that all wires are of equal length $L=L_{nm}$ and that all the junctions are of equal width $W=W_{nm}$. Under these conventions the array will be translationally invariant with respect to a unit cell consisting of an adjacent pair of a black and a white junction. The remaining free parameters are the 
superconducting 
phases $\phi_i$ on each wire $i$. Even if each wire is deposited on top of the same superconductor, the way the superconductivity is induced to the wires allows them to have different phases depending on their relative orientations. Following Ref. \onlinecite{Alicea2011} we adopt a convention that the phase difference will be directly proportional to the relative geometric angle between two wires. That is, if two wires meet at angle $\theta$ at the junction, then the relative superconducting phase is $\delta\phi_{ij}=\theta/2$. Thus all the parameters, except for the global parameters $\Delta$ and $\mu$, are fixed by the array geometry.

Our aim is to study the collective topological phases that can emerge in $N$-junction arrays for different array geometries. Before doing so, we will make a small detour and review the general spectral structure of Kitaev spin models. We will show that sectors of these models will also be described by Majorana tight-binding models that can be realized as the low energy theories of suitably constructed wire networks.

\section{Kitaev spin models} \label{sect:Kitaev}

Kitaev spin models are exactly solvable spin models defined on two dimensional lattices with trivalent vertices. The original model was defined on a honeycomb lattice \cite{Kitaev2006}, but the generalizations to other lattice geometries are straightforward \cite{Yao2007,Yang2007,Kells2011,Whitsitt2012}. The trivalent lattice geometry allows the links to be labelled as $x$-, $y$- and $z$-links such that one of each type will meet at every vertex. The Hamiltonian can be written as
\be \label{HKitaevSpin}
  H = \sum_{\alpha = x,y,z} \sum_{(i,j) \in \alpha-\textrm{link}} J_\alpha \sigma^{\alpha}_i \sigma^{\alpha}_j,
\ee
where $J_\alpha$ are the coupling strengths and $\sigma^{\alpha}_i$ are Pauli matrices acting on the sites $i$ of the lattice when $(i,j)$ is an $\alpha$-link. The key property underlying the exact solvability of all these models, regardless of the lattice geometry, is the presence of a local symmetry operator $\hat{W}_p$ on every plaquette $p$ of the lattice. These plaquette symmetries enable one to restrict to a particular sector $W=\{ W_p \}$ of the model labelled by the pattern of the local symmetry operator eigenvalues $W_p$. Their possible values depend on the lattice geometry. For plaquettes with an even number of links the eigenvalues are $W_p=\pm 1$, whereas for odd plaquettes they are given by $W_p=\pm i$. Complex eigenvalues imply that systems with odd plaquettes can spontaneously break time-reversal symmetry\cite{Yao2007}, while to break it in systems with only even plaquettes requires additional three spin interactions\cite{Kitaev2006,Kells2011}. Breaking time-reversal symmetry is of interest, 
since  only then can the system support topologically ordered phases with non-zero Chern numbers, i.e. ones that can support chiral Abelian (even Chern numbers) or non-Abelian (odd Chern numbers) anyons\cite{Kitaev2006}.

In each sector $W$ the spin problem can be mapped to a tight-binding problem of free Majorana fermions on the same lattice. Following the mapping introduced by Kitaev\cite{Kitaev2006}, the Hamiltonian takes the form
\be \label{HKitaevMaj}
  H_{W(u)} = i \sum_{\alpha = x,y,z} \sum_{(i,j) \in \alpha-\textrm{link}} J_\alpha u_{ij} \gamma_i \gamma_j,
\ee
where the Majorana operators $\gamma_i^\dagger=\gamma_i$ satisfy $\{ \gamma_i,\gamma_j \} = 2\delta_{ij}$ and $u_{ij}=\pm 1$ are local $Z_2$ gauge variables in a fixed gauge. They encode the sector through 
\be \label{W}
  W_p = -i^{|p|} \prod_{(i,j) \in p} u_{ij},
\ee
where $|p|$ is the number of links forming the plaquette $p$. In agreement with $u_{ij}$ being gauge variables, the spectrum depends only on the sector $W$, even if there are many configurations $u=\{ u_{ij} \}$ giving rise to the same $W(u)$ (we refer to Ref. \onlinecite{Kitaev2006} and \onlinecite{Pedrocchi2011} for more details). The plaquette operator expectation values \rf{W} can thus be viewed as expectation values of gauge invariant Wilson loop operators, which gives the following interpretation to their eigenvalues: The eigenvalue $W_p=\pm i, -1$ correspond to having a $\pm\pi/2$ or $\pi$-flux vortex on plaquette $p$, respectively, while $W_p=1$ denotes absence of one. Based on this we will refer to the sectors $W$ of Kitaev spin models as \emph{vortex sectors}. The Hamiltonian \rf{HKitaevMaj} is always quadratic in the Majorana fermion operators and thus readily diagonalized for arbitrary vortex sectors\cite{Feng2007,Chen2008,Lahtinen2008,Kells2009,Kells2010}.

The connection between the low-energy theories of topological wire networks and Kitaev spin models is then provided by the simple observation: \emph{If the array is constructed such that the wire ends coincide with the sites of a trivalent lattice, then the low-energy tight-binding model \rf{HN} will always realize some parts of the phase diagram of some vortex sector (in a fixed gauge) of the corresponding Kitaev spin model}. This observation enables one to immediately translate much what is know about the phase diagrams and stability of topological phases in Kitaev spin models into the wire network setting.  In the next section we will study this correspondence in detail using a particular example, namely that of the Yao-Kivelson (Y-K) variant\cite{Yao2007} that is realized as an $N=3$ junction array. Before doing so, we will briefly review what is known about the properties of the vortices in Kitaev spin models as they will have counterparts also in wire arrays.

\subsection{Vortices in Kitaev spin models}

The properties of isolated $\pi$-flux vortices ($W_p=-1$ eigenvalues on plaquettes far away from each other) depend on the topological phase the system is in. These can be characterized by the Chern number $\nu$, which directly gives the nature of the vortices\cite{Kitaev2006}: In  $\nu=0$ phases the vortices behave as achiral Toric Code anyons, in even $|\nu|$ phases they behave like chiral Abelian anyons and in odd $|\nu|$ phases they bind isolated Majorana modes and thus behave as non-Abelian anyons. While these properties are universal, the conditions under which a particular phase emerges depends on the particular variant of the Kitaev spin models.

Since the vortices correspond to symmetries of the Hamiltonian, they are static excitations. Their properties, depending on the Chern number $\nu$, are encoded in the low-energy part of the energy spectrum of the corresponding vortex sector. In the $\nu=0$ phases the vortex properties can be obtained analytically \cite{Kitaev2006,Schmidt2007,Kells2008}, but in the other phases this has to be done numerically by simulating vortex transport.\cite{Lahtinen2009} This has been explicitly studied in the $|\nu|=1$ phase of the original honeycomb model, where both the topological degeneracy\cite{Lahtinen2011} and the braid statistics\cite{Lahtinen2009,Bolukbasi2012} associated with the Majorana binding vortices has been verified. 

The key insight behind these studies is the observation that the vortex sector can be effectively changed by locally tuning the couplings $J_\alpha$. As one can see from \rf{HKitaevMaj}, the gauge variable $u_{ij}$ on link $(i,j)$ can be viewed as the sign of the corresponding local coupling $J_{ij}$ (or the intra-wire coupling $J'$). Thus from the point of view of the Hamiltonian, tuning adiabatically $J \to -J$  (or $J'$ to $-J'$) will interpolate between the spectra of two distinct vortex sectors that differ by the plaquette operator eigenvalues \rf{W} that depend on this link. This effectively amounts to creating/annihilating a vortex pair or transporting a vortex between adjacent plaquettes\cite{Lahtinen2011}. We will employ this same insight below to understand microscopic fluctuations in wire arrays in terms of vortices in the collective wire array states.
\begin{figure}
\includegraphics[width=.4\textwidth,height=0.40\textwidth]{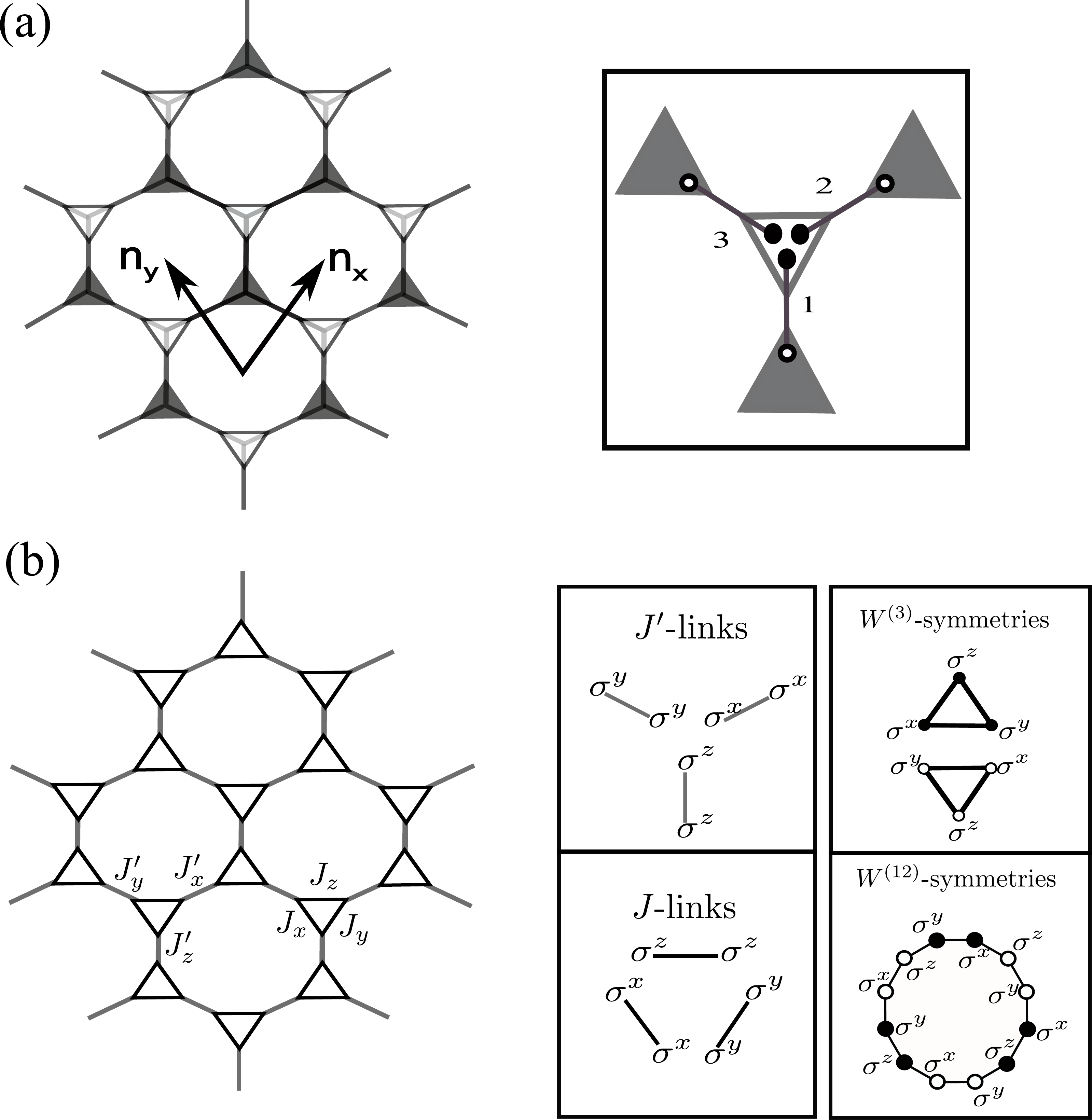}
       \caption{\label{fig:YK}  (a) The 3-junction nanowire array. Majorana states $\gamma_{b(w)}$ are denoted with black (white) circles. (b) The Yao-Kivelson variant of Kitaev spin models on the decorated honeycomb lattice. The couplings $J_\alpha$ span the triangular plaquettes, while the couplings $J_\alpha'$ connect them. } 
\end{figure}

\section{A $3$-junction network and the Yao-Kivelson model}
\label{sect:YK}
In this section we study in detail the correspondence between a 3-junction network and the Yao-Kivelson (Y-K) variant of Kitaev spin models on a decorated honeycomb lattice. First we will review the phase diagram of the Y-K model. Then we study which parts of it are realized in the wire array and show that phases with Chern numbers $|\nu|>1$ can be realized when the couplings are staggered in way that corresponds to an effective vortex lattices.

\subsection{The phase diagram of the Y-K model}

The Y-K variant of the Kitaev spin models \cite{Yao2007} is defined on a decorated honeycomb lattice that consists of both triangular and dodecagonal plaquettes, as illustrated in Fig.~\ref{fig:YK}.  We denote the corresponding plaquette operators describing the vortex sectors as $W^{(3)}=\pm i$ and $W^{(12)}=\pm 1$, respectively.  Fig.~\ref{fig:YK} also shows that the spin couplings of the Hamiltonian \rf{HKitaevSpin} on this lattice can partitioned into two sets: the couplings $J_\alpha$ act only on the links adjacent to the dodecahedral plaquettes, while the couplings $J_\alpha'$ are adjacent to both types of plaquettes with the triangular plaquettes consisting only of them. 

The ground state of the model is known to reside in a vortex sector where $W^{(3)}=\pm i$ uniformly on all triangular plaquettes and $W^{(12)}=1$ on all dodecagonal plaquettes. The phase diagram of this sector has been studied in several works \cite{Yao2007,Dusuel2008,Kells2010,Shi2010}. Defining $R= \sqrt{J_{x}^2+J_{y}^2+J_{z}^2}$ and $J'=J_x'=J_y'=J_z'$, the phase diagrams has the two distinct phases: For $R<J'$ the system is in a gapped $\nu=0$ phase that supports Abelian toric code anyons. For $R>J'$ the system is in a non-Abelian phase characterized by Chern number $\nu=\pm 1$, with the sign depending on the $W^{(3)}=\pm i$ sector. This phase can be mapped perturbatively to the non-abelian B-phase of the orignal Kitaev model \cite{Dusuel2008}, which in turn can be related to the weak $p+ip$ superconducting phase. \cite{Read2000} As described above, in this phase the $\pi$-flux vortices  ($W_p=-1$ eigenvalues) bind Majorana modes and  behave as non-Abelian Ising anyons.  Note also that when $R >2J'$ is satisfied, it is possible to consider non-uniform couplings $J_\alpha$ and $J_\alpha'$ for which a distinct $\nu=0$ phases can be obtained.\cite{Shi2010} However, our interest will mainly be be on the  phases emerging for the uniform couplings $J$ and $J'$.

\subsection{The 3-junction network and the Y-K model}

As illustrated in Fig.\ref{fig:YK},  the tight-binding model \rf{HN} for $N=3$ junction network is of the Y-K Hamiltonian form \rf{HKitaevMaj} where $u_{J'}  =  \textrm{sign}(J')$, $ u_J = \textrm{sign}(J)$ are the effective gauge fixed variables on the links of type $J'$ and $J$, respectively. Since we assumed $k_F$ to be equal in all wires and their lengths to be fixed to $L$, the $u_{J'}$ will be uniform across the array. The $u_J$ will also be fixed by the array geometry that fixes the relative superconducting phases. However, unless all the angles $\theta_{ij}$ are equal, not all the $u_J$ in the same junction have to be the same. Every junction will have the same pattern of couplings though, which implies that all types of links will appear twice in the effective plaquette operators. The dodecagonal plaquettes will then always take the value $W^{(12)}=u_{J'}^6 u_J^6 =1$, while the triangular plaquettes will have $W^{(3)} = i u_J^3=\pm i$ depending on the orientation $\theta$. Thus the ground state of the wire array maps into the ground state sector of the Y-K model, with the ground state coinciding with either of the two time-reversed ground states depending on the sign of $u_J$. 

\begin{figure}
\includegraphics[width=.35\textwidth,height=0.3\textwidth]{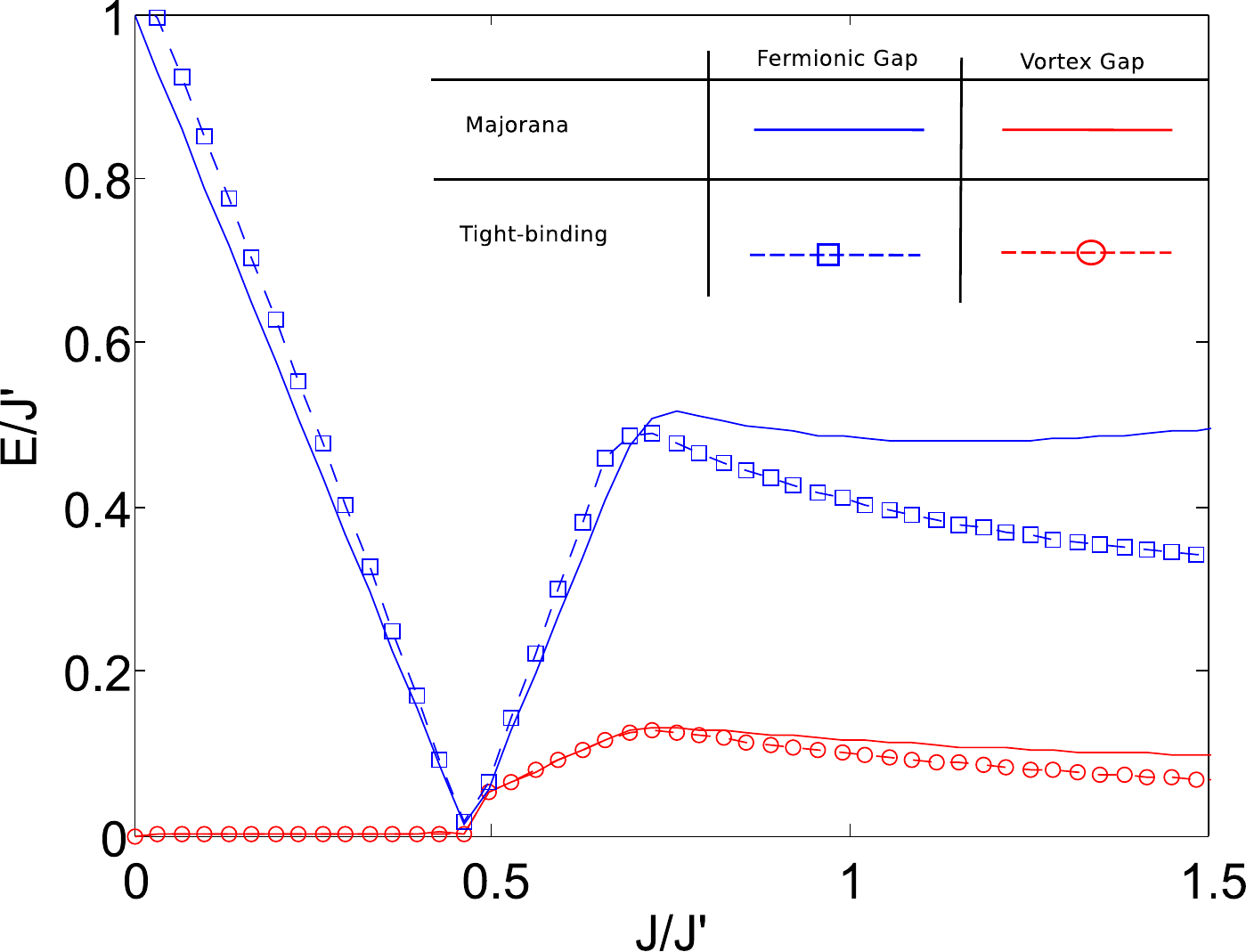}
       \caption{ \label{fig:FVgap} The fermion gap in the vortex-free sector (squares) and vortex gap (circles) calculated from the effective Majorana model \rf{HN} with derived couplings \rf{ZJ} (solid line) and from the full microscopic array model (dashed lines) presented in Appendix \ref{sect:App_micromodel}. By vortex gap we mean the ground state energy difference between the vortex-free sector and the a sector with two neighbouring vortices. The full microscopic model consists of 48 wires each of length $L=20$ (19 sites with lattice spacing $a=1$). The used parameters are $t=1,\mu=-1$ and $\Delta=0.15$ that correspond to a $L/\xi=1.5$ and $L/\lambda_F \approx 3$. The inter-wire tunnel couplings that model the Josephson couplings span $\tau \in [0,0.6]$ and the superconducting phases are taken equal, i.e. corresponding to $\beta=1/3$. }
\end{figure}

Thus we can immediately predict the form of the phase diagram as the function of $J/J'$, as shown in Fig. \ref{fig:FVgap}. When $R<J'$, which is in general satisfied for $W/\xi_J \ll L/\xi$, the system is in a state characterized by Chern number $\nu=0$, suggesting the Majoranas would form a collective Abelian state that would support Toric Code -type anyons. However, one should keep in mind that this phase is adiabatically connected to the limit of completely decoupled wires ($J_\alpha \to 0$), where the Majorana modes are isolated from each other. Thus there is no hybridization in the sense of undergoing a phase transition. The degeneracy in this decoupled wire limit is only lifted in a manner that gives rise to a Hilbert space that coincides with that of the $\nu=0$ phase emerging in the Y-K model. If one were to operate the wire array as topological quantum computer, it is in this $\nu=0$ phase where the system should be prepared and where it should remain at all times. 

This contrasts with the phase in the $R > J'$ regime, i.e. when both Josephson and the wire tunnelling couplings are of comparable strength. There the Majoranas hybridize and form an extended collective state across the whole array, which is characterized by Chern number $\nu=\pm 1$. Isolated Majorana modes at the wire ends are no longer localized low-energy excitation of the array. Had they been used for quantum computation, a transition to this phase would imply that some, but not necessarily all, \cite{Kells2013} encoded information would be lost. This phase still supports localized Majorana modes, but they appear now as collective modes centred at those dodecagonal plaquettes with $W^{(12)}=-1$.

\begin{figure}
\includegraphics[width=.45\textwidth,height=0.28\textwidth]{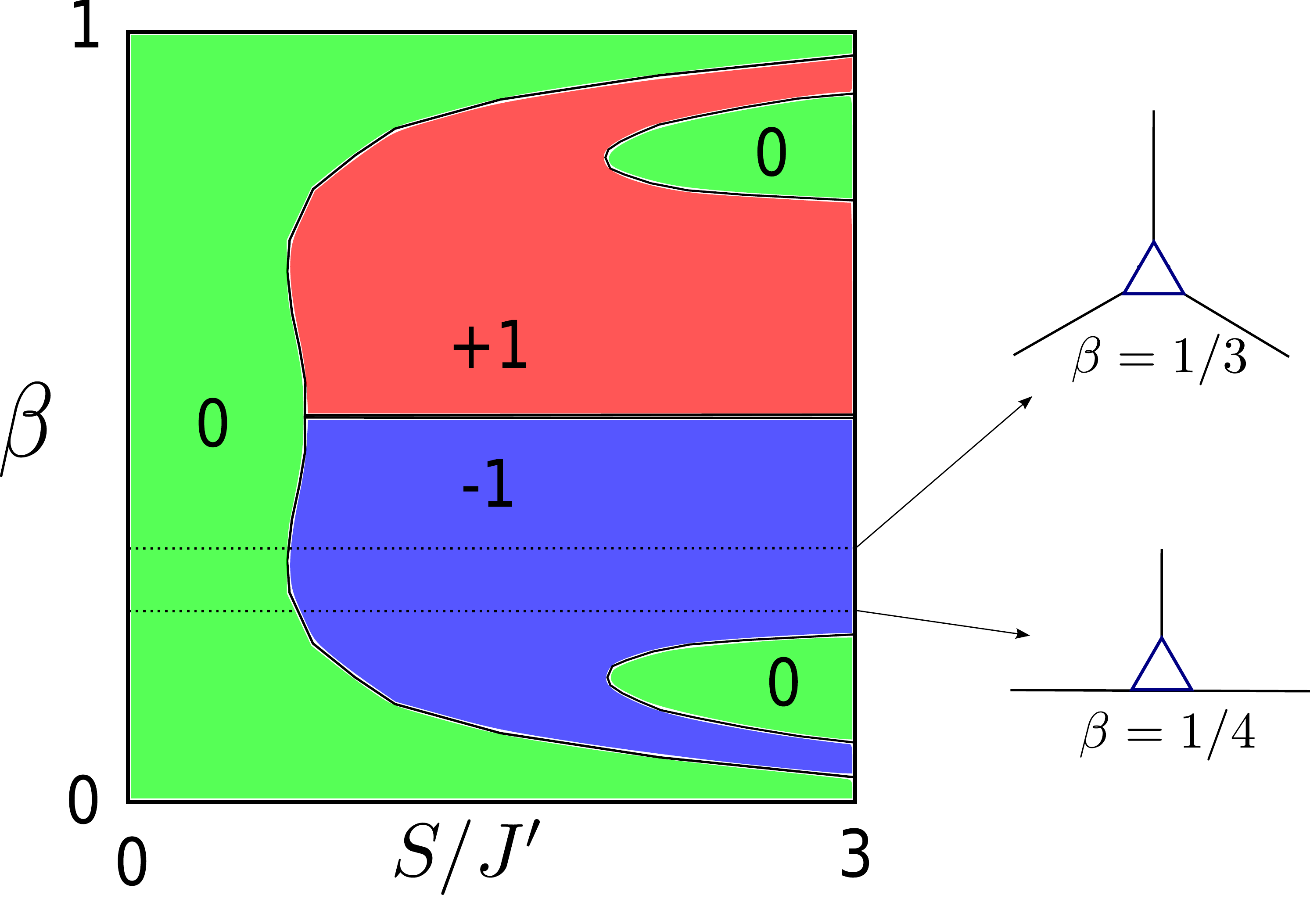}
       \caption{ \label{fig:N3_beta} Phase diagram of the 3-junction wire array as a function of the Josephson couplings (here $S = J / \sin \delta \phi$) and the array deformation parameter $\beta$ (which gives the relative superconducting phases $\phi_n = 2 \pi (n-1) \beta$). This parametrization enables us to consider the pure tunnelling effects separately from the array geometry that, under our simplifying assumptions, gives the superconducting phases. The $\nu=0$ phases in the $S/J' \ll 1$ regime correspond to parameter ranges where the wire end Majoranas do no hybridize, whereas in the $\nu=\pm1$ phases they form an extended collective state. The $\nu=0$ phases inside these phases emerge when $R>2J'$ is satisfied with the Josephson couplings $J_\alpha$ being unequal, as studied in Ref. (\onlinecite{Dusuel2008}). On the right we illustrate the uniform Y-K and brick wall array geometries that are obtained for $\beta=1/3$ and $1/4$, respectively.      
  }
\end{figure}

Having established that a wire array where all the wires meet at the same angle at each junction realizes the Y-K model, we can ask what happens if we deform the array by allowing the wires to meet at different angles. To study this systematically, we parametrize the three relative angles at a junction by $\theta_n= 2 \pi (n-1) \beta$. For $\beta=1/3$ one recovers the rotationally symmetric Y-K model, while $\beta \neq 1/3$ implies that only two out of the three $J_{\alpha}$ effective tunnelling couplings will now be equal. Fig. \ref{fig:N3_beta} shows that $\beta$ deformations have in general only a small effect in the phase diagram around the Y-K model. If Josephson couplings are larger than the tunnelling couplings $J$, we find that that another $\nu=0$ phase can open inside the hybridized $\nu=\pm 1$ phases. This phase is adiabatically connected to the phase that is known to emerge in the Y-K model when $R > 2 J'$ is satisfied while the $J_\alpha$ are unequal\cite{Dusuel2008}. Finally, we note that the time-reversal symmetry between $0 < \beta < 1/2$ and  $1/2 < \beta < 1$ follows from one the $u_J$'s changing sign at $\beta=1/2$ which means that there is transition between time-reversed phases belonging to sectors $W^{(3)}=+i$ and $W^{(3)}=-i$, respectively.

\subsubsection{Comparison to a full microscopic model for the wire array}

When deriving the Majorana model for the wire array we assumed the Josephson couplings to be perturbations to a system of decoupled wires. Thus one expects the model to provide an accurate description of the system in the small $J/J'$ limit. To quantitatively study the accuracy of the effective Majorana model \rf{HN}, we compare the energy gaps calculated from it to those calculated from a full microscopic model for the wire array, i.e. to one where we do not assume a priori the existence of Majorana end states. Such a tight-binding model for an array of $p$-wave wires is presented in Appendix \ref{sect:App_micromodel}. 

Figure \ref{fig:FVgap} shows that the energy gaps for both fermionic and vortex excitations are in excellent agreement until about $J/J' \approx 1$. For larger relative values the Majorana model starts slightly to overestimate their magnitudes. Still, all the phases remain robust, which suggests that the qualitative description provided by the Majorana model is correct beyond the limit of treating the Josephson couplings as small perturbations. The derived Majorana model \rf{HN} thus provides for $J/J'<1$ an accurate quantitative description of the low-energy physics of the topological wire array, with qualitative features captured also for  $J/J'>1$.

\subsection{Effective vortices in nanowire arrays}

As we discussed above, the Kitaev spin models support vortex excitations whose presence in the $|\nu|>0$ phases could be related to sign flips in the spin couplings. Their counterpart in the wire arrays are then the sign flips in the tunnelling and Josephson couplings \rf{ZJ}. These can change signs either through variation in the array parameters due to uncertainty in the construction of the array or due to thermal and quantum fluctuations.

In the first case the tunnelling couplings $J'$ can change sign if the length $L$ of a wires varies locally such that the $k_FL$ varies at the scale corresponding to half of the Fermi wavelength. On the other hand, $J$ can change sign due to local deformations of the array that result in unequal $\beta$'s on different junctions. Thus imperfections in the array construction can result it realizing some other vortex sector of the Y-K model than the vortex-free sector that contains the ground state. One can also turn this perspective around: By intentionally creating local geometric deformations of the array one can create states with static patterns of vortices. Below we will show that this insight can be used to create effective vortex lattices that in principle enable $|\nu|>1$ Chern number phases to be realized. 

The other general way for the couplings to flip signs is through changes of the chemical potential and/or the superconducting phase due to fluctuations in the electron density of the underlying $s$-wave superconductor. While the exact likelihood of these effects will depend on the microscopic realization of the array, we can make qualitative statements about their relevance and consequences based on the vortex properties known from the Kitaev spin models. As shown in Fig. \ref{fig:FVgap}, the vortices are massive excitations in the $|\nu|=1$ phases, while in the $\nu=0$ phase where they are essentially gapless (their mass scales as $(J/J')^6$). This means that in the first case the collective state energetically suppresses fluctuations that could excite them, whereas in the latter case they are essentially free to be created and transported around the array. We discuss in Section \ref{sect:Stability} the consequences of this to the stability of the array as an quantum computer.

In addition to local fluctuations the superconducting phase of a wire may also spontaneously change by $\phi_i \to \phi_i + 2\pi$, i.e. undergo a phase slip \cite{Pekker2013}.  In topological superconductor junctions the Josephson coupling depends on the half the phase difference, which means that under a phase slip all the $J$ couplings connecting to this wire will also change sign. However, due to the trivalence of the lattice, this will not change any of the effective plaquette operator eigenvalues and thus no effective vortices are excited. In the context of Kitaev spin models, such transformations would correspond to gauge transformations, because all physical observables of the system will remain unchanged. While two coupling configurations related by such transformations are physically distinct in the array, from the point of view of the collective state they are equivalent. Thus it seems that the ground state of a wire array not only realizes the Y-K model in a fixed gauge, but that phase slips also 
provide the counterpart of gauge fluctuations. 

Summarizing, vortices in the array can be created either as static defects in the array construction or as dynamical excitations due to local fluctuations of the chemical potential and the superconducting phase. Implications of both processes to the stability of the emerging phases in wire arrays are discussed in Section \ref{sect:Stability}.

\subsection{Higher Chern number phases from staggered couplings} 
\begin{figure}[t]
\begin{tabular}{c}
\includegraphics[width=.3\textwidth,height=0.17\textwidth]{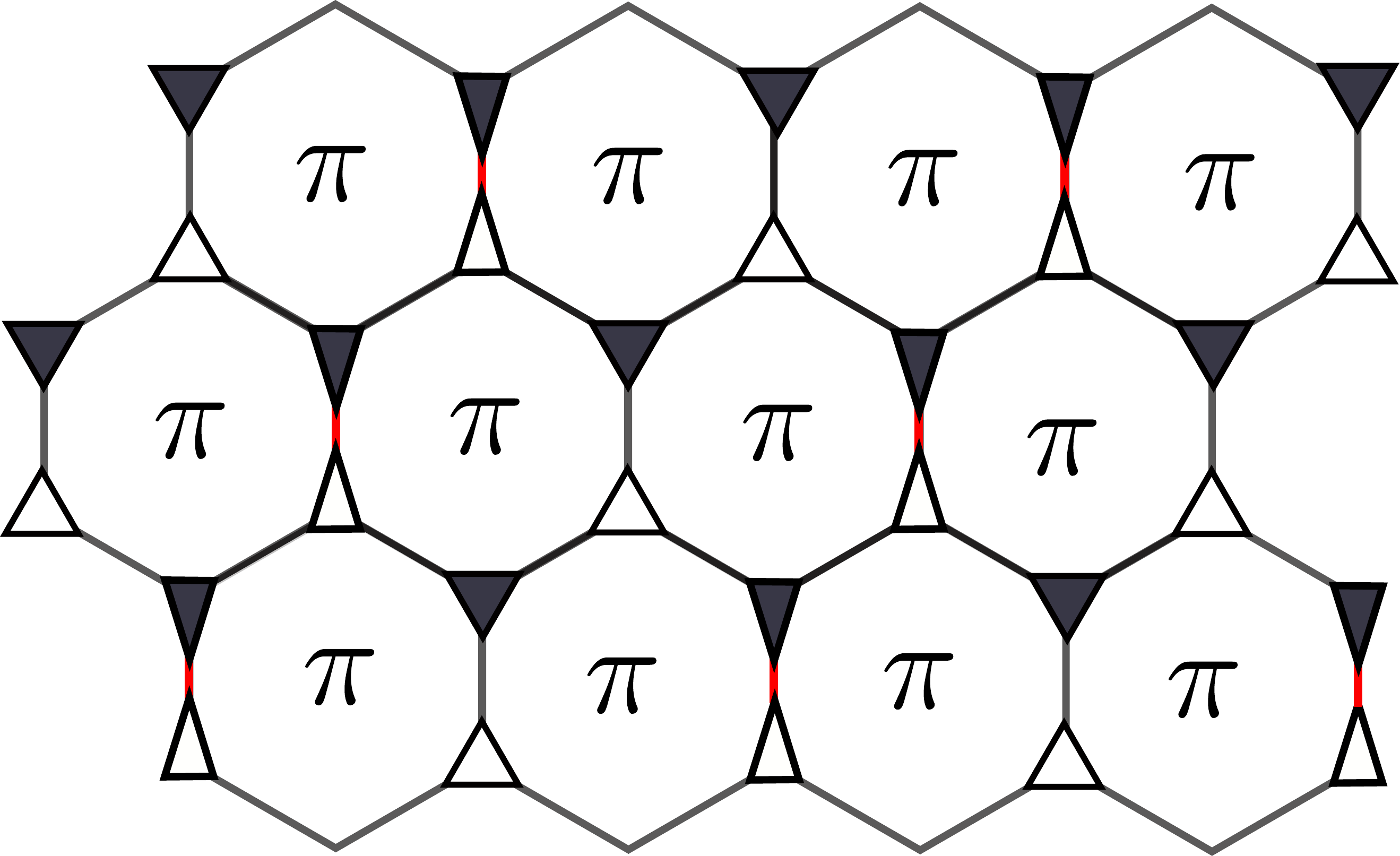} \\
(a) \\
\includegraphics[width=.35\textwidth,height=0.3\textwidth]{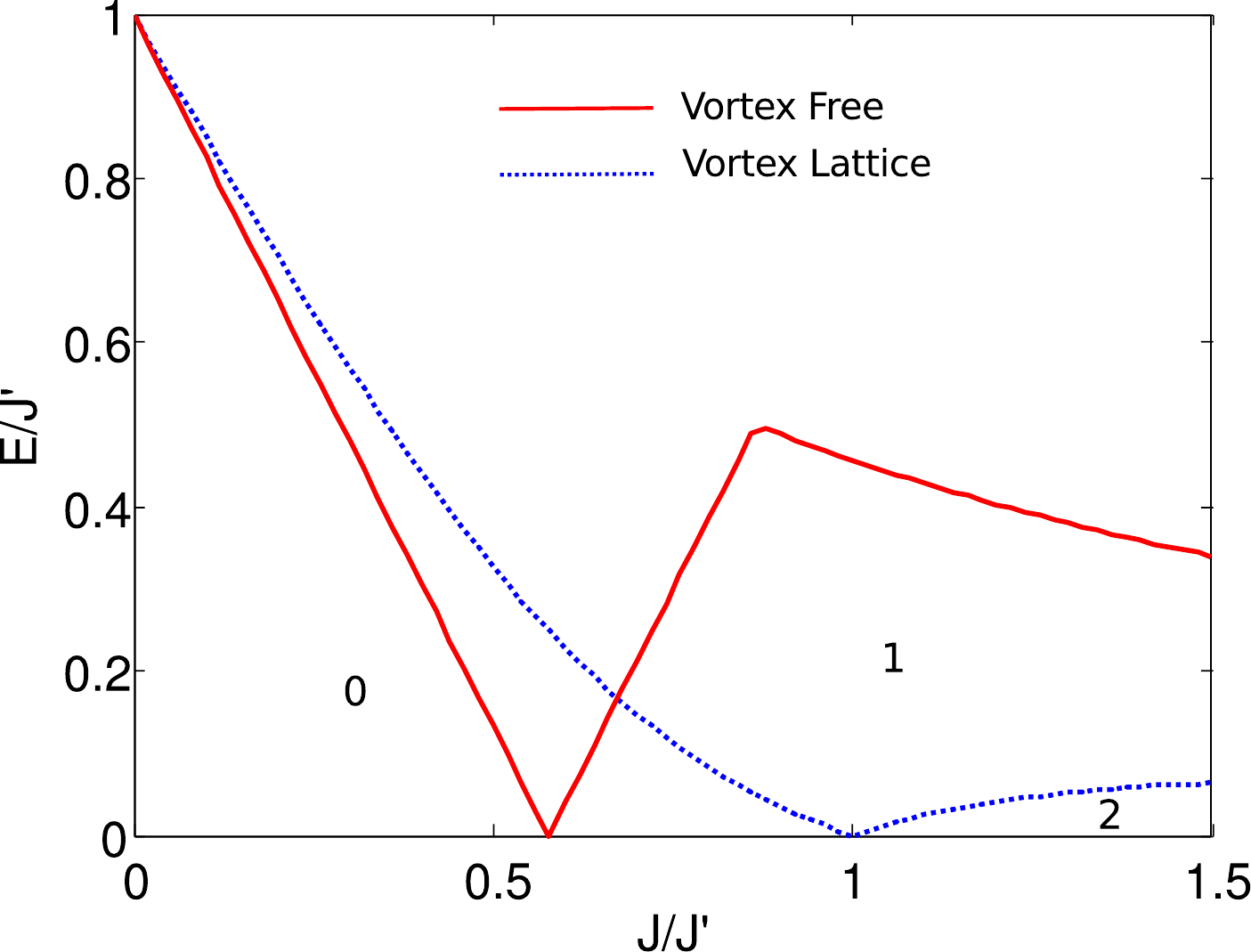} \\
(b)
\end{tabular}
\caption{ \label{fig:VL_array} The vortex lattices and higher Chern number phases in the Y-K model. (a) By alternating the lengths of the wires (or equivalently staggering the chemical potential), the tunnelling couplings can become sign staggered such that each thin black link has $J$ while every thick red link has $-J$. This corresponds to having a $\pi$-flux vortex ($W^{(12)}=-1$) on every dodecagonal plaquette. (b) In the presence of such staggering, we find that the gap closure moves to a larger value of $J/J'$ and the collective state is now characterized by $\nu=2$. Both features are consistent with the studies in the honeycomb model.\cite{Lahtinen2012}}      
\end{figure}
We have argued above that a uniform 3-junction array realizes the vortex-free sector of the Y-K model that contains the ground state across all sectors. Consistent with what is known about the phase diagram of the model, we only find phases characterized by Chern numbers $\nu=0$ or $\pm 1$. For the 3-junction array to realize other Chern number phases, one needs to effectively realize other vortex sectors. 

One way to do this is to introduce periodically sign staggered couplings that will correspond to uniform vortex lattices. This has been studied in the context of the original honeycomb model\cite{Lahtinen2012}, where different phases with Chern numbers $\nu=\pm 2, \pm 4$ can be realized depending on the spacing of the vortex super-lattice. To verify that this same mechanism works also in the Y-K model, we plot in Fig. \ref{fig:VL_array} the phase diagram when one has $W_p^{(12)}=-1$, i.e. a $\pi$-flux vortex, on every dodecagonal plaquette. Like in the honeycomb model these vortices form a triangular super-lattice and as predicted by the previous studies \cite{Lahtinen2010, Lahtinen2012}, we find that the $\nu=\pm 1$ phases are indeed replaced by $\nu=\pm 2$ phases. Based on this we postulate that also the $\nu=\pm 4$ phases are realizable in this way when the effective vortex lattices are sparser.

To induce the effective vortex lattice in the wire array, one needs to have signs of the tunnelling amplitudes \rf{ZJ} staggered in a suitable manner. To do this physically we need to relax our assumptions that all the wires are of equal length. For instance, one way to construct a vortex lattice with $W_p^{(12)}=-1$ on every dodecagonal plaquette is to have $u_{J'}$ alternate on the horizontal links of every row. As illustrated in Figure (\ref{fig:VL_array}), this could be achieved by allowing the lengths of adjacent wires to vary at the scale of the half the Fermi wavelength, i.e. break translational invariance at the level of the array construction. Alternatively, the same effect could be achieved by having the wires independently gated such that the chemical potential is staggered at the scale of $~ \pi^2/2m L^2$ to stagger $k_F$ in a corresponding manner. A third option is to allow different wire orientations in different junctions, which would cause the $u_J$'s to acquire the required staggering.

As the construction of the uniform arrays is likely to be challenging, we leave the analysis of the feasibility of realizing such staggered arrays for future work. Our motivation here is merely to point out that given sufficient experimental precision, there exists a straightforward recipe for constructing arrays supporting collective states with Chern numbers $|\nu|>1$.

\subsection{Higher Chern numbers in higher $N$-junction arrays}
An alternative method of achieving higher Chern numbers is to go to higher $N$-junction arrays. Like the $N=3$ array that maps to the Y-K variant of Kitaev models, the $N=4$ array maps into the so called Square-Octagon model, that is know to host phases with Chern numbers $0,\pm1,\pm2,\pm3,\pm4$ given that longer range tunnelling is sufficiently strong \cite{Yang2007,Kells2011}. These are naturally present in $N \geq 4$ arrays due to there being always more than just nearest neighbour tunnelling across each junction. However, as longer range couplings they also tend to be exponentially weaker in the junction width.

We have analysed in detail the $N=4$ array shown in Figure \ref{fig:arrays}(a) in Appendix \ref{App_Higher}. We find that the longer range tunnellings across the junction are insufficient to reach any other phases except those characterized by $\nu=0$ and $\pm 1$. On the other hand, for an array with alternating 6- and 3-junctions, illustrated in Figure \ref{fig:arrays}(b), we find that junction couplings of three different ranges are sufficient to open up robust phases with $|\nu|=2$. We believe that the full phase diagrams of the higher $N$ arrays can be very rich, but as their realizations are likely to be challenging, we again leave studying them to future work.

\begin{figure}[t]
\includegraphics[width=.40\textwidth,height=0.16\textwidth]{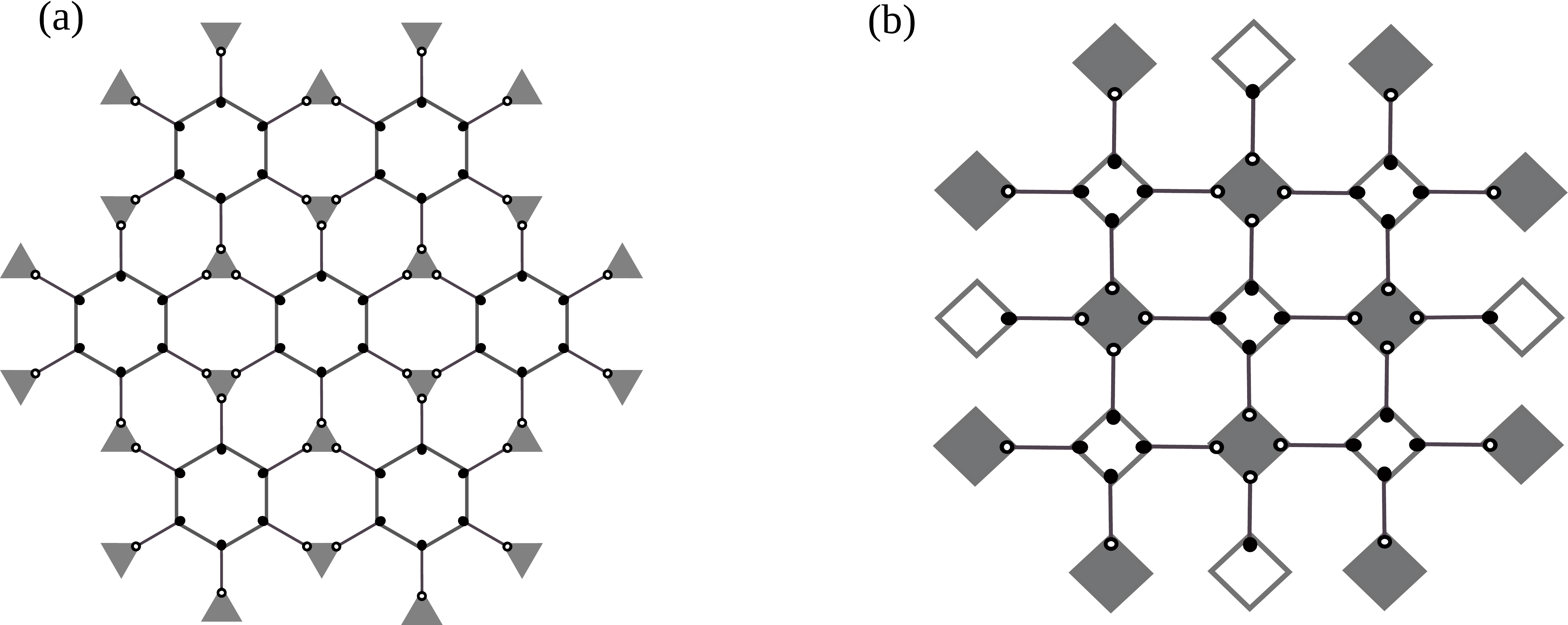}
       \caption{ Illustrations for (a) an array with alternating 6- and 3-junctions and (b) a 4-junction array. Majorana states $\gamma_{b(w)}$ are denoted with black (white) circles.}
 \label{fig:arrays}
\end{figure}

\section{Stability of the collective states in wire arrays}
\label{sect:Stability}
We have argued that wire arrays support collective phases characterized by different Chern numbers and that these phases are in one-to-one correspondence with those appearing in Kitaev spin models.  A natural question to ask is how stable these collective states are and how much about their stability can be inferred from the stability of the corresponding phases in the spin models. 

To this end we consider the wire array in the presence of local random electrostatic disorder. Formally this means that the chemical potential $\mu$ becomes a local random variable along each wire, which at the level the effective Majorana model translates to the tunneling couplings $J_\alpha'$ becoming local random variables. To study electrostatic disorder quantitatively, we model it as an additional Gaussian white noise potential along each wire with mean $\langle V(\vr) \rangle = 0$ and variance $\langle V(x) V(x') \rangle = \alpha \delta(x-x')$. Here $\alpha= v^2_F/l$, $l$ is the mean free path and $v_F = p_F/m$. As the phases of the array arise as collective states of the Majorana modes, an absolute upper bound for their stability can be inferred from the condition that each wire remains in the topological phase that supports the Majoranas. The effect of local random disorder on a single p-wave wire was studied in a number of works,\cite{Motrunich2001,Brouwer2011,DeGottardi2013} with the exact stability conditions depending on the microscopic details of the wire.\cite{Rieder2013,Brouwer2011b,Adagideli2013} For an ideal single-band wire we can use the general result of Ref. \onlinecite{Brouwer2011} where it is shown that the Majorana end states persist as long as $ \xi/l \lesssim 2$. We take this to be also the absolute upper bound of the wire array in the presence of local random disorder.

However, before the outright failure of individual wires, disorder may drive the wire array system collective state. For disorder that is not strong enough to drive individual wires out of the topological phase, we identify two distinct regimes based on the behavior of disordered Kitaev spin models\cite{Willans2010,Chua2011,Lahtinen2013}: (i) Weak tunneling disorder, when only the amplitudes of the tunneling amplitudes $J$ and $J'$ become random, and (ii) strong tunneling disorder, when they can also change signs.


\subsection{Stability of the collective states with odd Chern numbers}

Let us consider first the stability of the  $\nu=\pm 1$ collective states.  If disorder in the wire is weak enough, it only causes local amplitude randomness in the tunneling couplings $J$ and $J'$. This type of tunneling disorder has also been studied in the context of Kitaev spin models\cite{Willans2010,Chua2011,Lahtinen2013}. The result is that the energy gap of the collective state decreases monotonously with increasing disorder strength $\alpha$. All the qualitative properties of the phase remain invariant though and thus the phases are stable with respect to moderate disorder. We expect this result to apply also to wire arrays with one caveat. A decreasing energy gap implies a growing coherence length $\xi$, which in turn implies that $J/J'$ decreases. Assuming that everything else remains invariant, weak disorder can thus drive the system towards the $\nu=0$ phase. This can lead to a phase transition if the system is prepared in the $J/J'<1$ regime close to the phase transition, as shown in Figure \ref{fig:FVgap}.

Something more dramatic can occur for strong tunneling disorder, i.e. when the couplings $J'$ can also have random signs. This happens when disorder causes $k_F$ to vary locally at the scale of the inverse Fermi wavelength. We can estimate the required disorder strength by assuming that the Majorana overlap integral giving the coupling $J'$ depends cumulatively on $k_F(x)$ in the wire. In other words, we assume that $J' \sim \sin( \int_0^L k_F(x) dx) $,  where $k_F(x)=\sqrt{2 m (\mu -V(x))}$ is the local Fermi momentum. As $\alpha$ is still small compared to the average chemical potential, we can approximate the integral as
\be
\int_0^L k_F(x) dx  \approx k_F L + \frac{2}{v_F} \int_0^L V(x) dx . \nonumber
\ee
The last term has a zero mean and standard deviation $\sigma = 2 \sqrt{L/l}$. For sign flips to occur in the tunnelling amplitudes, as a general rule of thumb we then require that the standard deviation is of the order of the $\pi$-shift required to change $J' \to -J'$. This leads to the condition $L/l > \pi^2/4$. Unless the wires are very short ($L \approx  \xi$), this clearly is a more stringent condition than $\xi/l < 2$, i.e. sign disorder occurs before individual wires are driven out of the topological phase.  

As we have discussed above, sign flips are equivalent to creation of vortices and thus the onset of sign disorder in the Majorana tunneling can equivalently be viewed as an emergence of a random vortex lattice. In the $\nu=\pm1$ phases the vortices bind Majorana modes, which means that sign disorder gives rise to a random Majorana hopping problem defined on a dual lattice. This problem has been considered in Ref. (\onlinecite{Laumann2012}), where sufficient randomness of the signs is predicted to drive the system into a gapless thermal metal state. This mechanism has been shown to hold in the context of the honeycomb model\cite{Lahtinen2013} and thus it is expected to apply also in the variants of Kitaev spin models. Thus we predict that when $L/l \gtrsim \pi^2/4$ and $\xi/l \lesssim 2$, i.e. roughly when
\be
  \xi \lesssim 2l \lesssim L,
\ee
also the wire array can be driven into this disorder induced thermal metal state that is characterized by a logarithmically diverging density of states.\cite{Laumann2012} Note that the condition $L/l > \pi^2/4$ suggests that arrays with very long wires are more susceptible to disorder of this type. However, here one should keep in mind that this absolute limit corresponds $J/J' \ll 1$, where the array would be in the $\nu=0$ phase. There vortices will not bind Majorana modes and thus the thermal metal state can not emerge.
 
Finally, it should be noted that the emergence of the thermal metal state is based only on tunneling disorder in the low-energy Majorana model. Apart from electrostatic disorder, it could as well arise  due to randomness in the wire lengths, junction widths or relative angles at junctions, which all will always translate into tunneling disorder for the Majoranas. Thus qualitatively similar behavior can be expected also for these other types of disorder arising from imprecise construction of the array. Thus assuming that $\alpha$ also parametrizes uncertainty in the wire lengths or junction widths, we can take $L/l \gtrsim \pi^2/4$ also as a guideline for the required precision to construct robust collective states in the topological wire arrays. We also expect that local superconducting phase or thermal fluctuations can give rise to qualitatively similar effects. Small fluctuations will lead only to amplitude fluctuations of the $J$ couplings, while large fluctuations can also cause them to flip signs. Thus the thermal metal state may also emerge due to them.\cite{Bauer2013}

\subsection{Implications for the array as a quantum computer }

The nanowire arrays, when used as a topological quantum computational architecture, are operated in a regime where the Josephson couplings $J$ are much weaker than the intra-wire couplings $J'$\cite{Alicea2011,Sau2011,vanHeck2012,Hyart2013,Fulga2013}. In the Y-K model this regime corresponds to the $\nu=0$ phase that supports achiral Abelian (Toric Code) anyons, see Figure \ref{fig:FVgap} and Refs. \onlinecite{Yao2007,Dusuel2008}. This enables one to view some decoherence mechanisms as a proliferation of the low-energy vortex-like quasiparticles. 

To see this, lets consider the 3-junction wire array as a topological quantum computer, where the initialized computational space has all fermionic modes associated with the (nearly) decoupled wires un-occupied. As the Majoranas are braided, by locally controlling the amplitudes of the couplings $J'$ and $J$, the groundstate wavefunction undergoes a topologically protected evolution. To remain in the computational space at all times one must restrict to manipulations such that the Josephson couplings $J$ never change sign.\cite{Alicea2011,Sau2011}. As we have discussed earlier, we can recognize that this constraint is equivalent to demanding that no low-energy vortex excitations of the Toric-Code-like phase are excited. 

Accidental sign flips in $J$ can only come about through changes in the phase of superconducting order parameter.  In the Toric Code picture, such sign flips amount to spontaneous creation and propagation of the vortices. In real world realizations, these phase fluctuations are expected to be suppressed because the order parameter is inherited from a macroscopic superconductor. However, on the level of the p-wave toy model, the essential gaplessness of the low-energy vortices (their mass scales as $(J/J')^6$) means that sign flips of the Josephson couplings are not suppressed.
Using the perspective of  Ref. (\onlinecite{Pekker2013}) where $2\pi$ phase slips are a potential source of dephasing noise, we can then immediately understand the proliferation and propagation of the low-energy Toric Code vortices as the counterpart of decoherence in the nanowire setup, see Appendix \ref{sect:FVS} for more details.   

Although this picture of proliferating Toric-code vortices is complementary to previous work on stability, (see e.g. Ref. \onlinecite{Fulga2013} for an analysis of braiding in the presence of disorder), our hope is that this perspective will encourage diffenrent approaches to the problem of fault tolerance in wire arrays. Toric-Codes, as the archetypal topological quantum memory, has been the subject of much research concerning its stability under perturbations (see e.g. Ref. \onlinecite{Wootton2012} for a recent review). It would be interesting to study whether some of these results could also be translated and applied to quantum computating with topological nanowire arrays.

\section{Conclusions}
\label{sect:Conclusions}
 
We have explored the possiblity of constructing wire arrays of topological nanowires that would support collective states of Majorana modes. Our main result is that for appropriate geometries these arrays can realize the same physics as exactly solvable Kitaev spin models.\cite{Kitaev2006} This connection is based on both systems, while being microscopically distinct, admitting low-energy desciption in terms of the tight-binding model of Majorana modes. By explicitly considering the Yao-Kivelson variant of these spin models on a decorated honeycomb lattice,\cite{Yao2007} we showed that an array of 3-junctions (three nanowires meeting at each Josephson junction) could support collective states characterized by non-zero Chern numbers. These emerge when the fractional Josephson couplings describing the coupling of Majoranas between different wires could be made comparable to the coupling between the two Majoranas residing at the ends of the same wire. Finally, we applied results from disordered Kitaev spin models to argue for the stability for these phases in the presence of both local random disorder and quantum and thermal fluctuations.

To experimentally construct a nanowire array that would support such collective states, one needs to realize the two elementary building blocks: The $p$-wave nanowire\cite{Kitaev2001} with Majorana end states and a Josephson junction of such wires.\cite{Kwon2004} Experiments on the first have already been carried out\cite{Mourik2012,Das2012,Churchill2013} and they give evidence for the existence of Majorana end states. Considering this recent rapid progress, it is concievable that also the fractional Josephson junction effect could be observed in a laboratory in the near future. Beyond these two elementary building blocks, there is no fundamental obstacle for the construction of nanowire arrays, with the robustness of the collective states being predominantly determined by the precision of the array construction. The detection of these states is a topic that we did not touch in the present work, but one would expect that the different states would have signatures in the transport properties across the array.\cite{Grosfeld2006,Medina2013}

Taking an optimistic view on the required experimental advances to construct nanowire arrays, there exists interesting many-body physics associated with Kitaev spin models that the wire array could be used to probe. Due to the richness of their phase diagrams\cite{Kitaev2006,Yao2007,Yang2007,Tikhonov2010,Kells2011,Whitsitt2012}, the immediate interest would be on topological phase transitions. We outlined also the conditions where wire arrays could be made to undergo more exotic transitions, such as a disorder induced transition to a metallic state\cite{Laumann2012} or a nucleation transition due to the presence of a vortex crystal.\cite{Lahtinen2012} Furthermore, if local control over the array parameters can be executed with sufficient accuracy, one could even entertain the possibility of using them to test non-Abelian braiding statistics \cite{Lahtinen2009, Bolukbasi2012}.  

Finally, let us conclude by summarizing the implications of our results for the quantum computing with nanowire arrays \cite{Alicea2011,Sau2011,vanHeck2012,Halperin2012,Fulga2013} that we used initially to motivate our work. As the formation of the $\nu=\pm 1$ collective states constitutes a signifianct source of decoherence, the most obvious impact is the understanding of how to avoid such scenario. Since this occurs in general only when the Josephson couplings are comparable to the intra-wire couplings, our results show that this can be avoided by keeping the junctions wide. The second implication of our results is that the regime where the array would be operated as the quantum computer corresponds to the $\nu=0$ phase in the corresponding spin model. This implies that the computational space coincides with the Hilbert space that supports Abelian (Toric Code) anyons and that dephasing decoherence in the array could equivalently be viewed as the creation and propagation of these anyonic quasiparticle excitations. The stability of the Toric Code systems, as the archetypal topological quantum memory, has been the subject of much research (see Ref. \onlinecite{Wootton2012} for a recent review). It would be interesting to study whether some of the stabilization schemes, such as local random potentials\cite{Wootton2011} or couplings to external baths\cite{Pedrocchi2011-2} could also be translated to increase the fault-tolerance of topological nanowire arrays.

\section*{Acknowledgements}

We would like to acknowledge the financial support of Science Foundation Ireland under the Principal Investigator Award 10/IN.1/I3013 (GK and JV), the Alexander von Humbolt Foundation (GK), the Dutch Science Foundation NWO/FOM (VL). GK would like to thank Dganit Meidan, Maresa Rieder and Falko Pientka for useful discussions.

\appendix

\section{A full microscopic model for the wire array}
\label{sect:App_micromodel}

In this Appendix we present a full microscopic model for an $3$-junction array of topological nanowires. We show that its predictions are in agreement with the effective Majorana model \rf{HN}, which justifies the use of the latter to study the nanowire arrays.

Instead of the continuum solution \rf{H1Dcont}, we model a single $p$-wave nanowire as a 1D lattice model on $N_L$ sites first introduced by Kitaev\cite{Kitaev2001}. The Hamiltonian for a single wire $n$ is given by
\begin{eqnarray} \label{Hmicro}
  H^n & = & - \mu' \sum_{l=1}^{N_L} c^{\dagger}_{l,n} c^{\phantom \dagger}_{l,n} \\
 \ & \ & - \sum_{l=1}^{N_L-1} \left( t c^{\dagger}_{l,n} c^{\phantom \dagger}_{l+1,n} +|\Delta_n| e^{i\phi_n} c^{\dagger}_{l,n} c^{\dagger}_{l+1,n} + \mbox{h.c.} \right), \nonumber
 \label{eq:H0} 
\end{eqnarray}
 where $\mu'= \mu-2t$, $t$ the hopping energy, $|\Delta|$ the magnitude of the pairing potential and $\phi_n$ the superconducting phase. These are related to the continuum parameters through $t=1/(2 m a^2)$  and $\Delta_n = \Delta_n/(2a)$, where $a$ is the lattice constant and the wire length $L=(N_L+1)a$. We assume that the overall sign of superconducting phase on each wire is defined with respect to start of the wire at $l=1$. 

Let us adopt a convention that all wire end sites labelled by $l=1$ come together in junctions of one type (say, white junctions in the main text) and the ones labelled by $l=L$ come to together in others (black junctions). To write down a microscopic Hamiltonian for an $N$-junction array, we then couple individual wires in each junction using the tunnelling terms between their ends
\be
H^\tau_i = \sum_{n<m}^N \tau_{nm} \left( c^{\dagger}_{l,n} c^{\phantom \dagger}_{l,m} +\mbox{h.c.} \right), \quad l=1,L.
\ee
The full microscopic model for a periodic wire array consisting of $N_w$ wires connected through terms like $H^\tau$. The tunnelling amplitude $\tau_{nm}$ represents the transmission through the barrier of height $V_0$ that we use to model the junction.  However, as it is written here it resembles a kinetic hopping term. For the purpose of deriving an effective Majorana hopping model it is extremely convenient form from which we can perturbatively calculate the Josephson tunnel coupling between different wires.

To compare the prediction by this full microscopic model to that of the effective Majorana model, we need to know how the microscopic parameters of \rf{Hmicro} relate to the effective Majorana couplings \rf{ZJ}. These can be obtained for solving for the Majorana end states in \rf{Hmicro}, which on the lattice are given by
\bea 
\gamma_b & = & \frac{1}{\sqrt{2 \mathcal{N}}} \sum_{l=1}^{N_L} \left[  e^{~i\phi/2} c_l^\dagger + e^{-i\phi/2} c_l \right] u(l), \\
\gamma_w & = & \frac{i}{\sqrt{2 \mathcal{N}}} \sum_{l=1}^{N_L} \left[ e^{~i\phi/2} c_l^\dagger - e^{-i\phi/2} c_l \right] u(N_L+1-l), \nonumber
\eea
where now $u(l) =R^l \sin( \theta l)$ with
\be
R = \sqrt{\frac{t-|\Delta|}{t+|\Delta|}}, \quad \theta =\cos^{-1} (\frac{-\mu+2t }{2 \sqrt{t^2-|\Delta|^2}}). \nonumber
\ee
Like in the main text, we can expanding the full microscopic Hamiltonian \rf{Hmicro} in the $\gamma_{b/w}$ basis. This gives for the nearest neighbour tunnelling couplings
\bea
J'_{nn} & = & i \frac{(t+|\Delta|) u(N_L+1) u(1) }{\mathcal{N}}, \\
J_{nm} & = & i \frac{\tau_{nm} u(1)^2 }{\mathcal{N}} \sin \delta \phi_{nm}. 
\eea
One can also obtain an expression for next-nearest neighbouring coefficients
\be
K_{nm}= i \frac{\tau_{nm} u(N_L) u(1) } {\mathcal{N}} \cos \delta \phi_{nm},
\ee
which describes tunnelling between different wires and different junctions. 

As we show in Fig. \ref{fig:FVgap}, the fermion and vortex gaps as calculated from this model are in excellent agreement with the the ones calculated from the effective Majorana model. The inclusion of the next nearest neighbour $K$-terms improves quantitative agreement, but as second order terms $J$ an $J'$ they are in general an order of magnitude smaller and thus they can be safely ignored. We have verified that in the $3$ and $4$-junction cases they are too weak to drive the system into a higher Chern number phase that these systems could in principle support.\cite{Kells2011,Whitsitt2012}

\section{Summary of results for higher $N$-junctions}
\label{App_Higher}

 \begin{figure}
\includegraphics[width=.35\textwidth,height=0.3\textwidth]{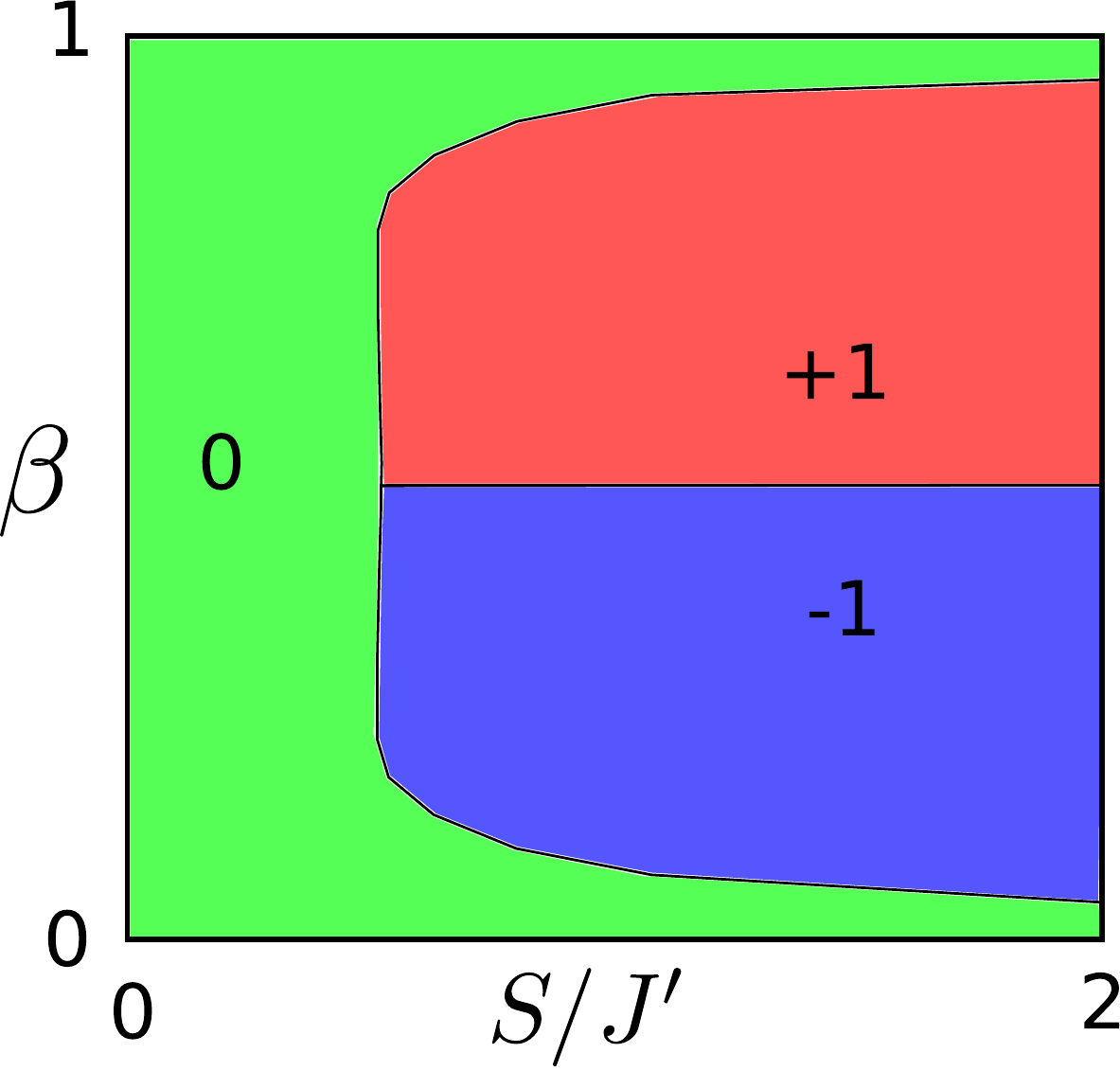}
       \caption{ The phase diagram for the 4-junction array with $S^{(1)}=S^{(2)}=S$.}
       \label{fig:N4_beta}
\end{figure}

Here we summarize the results for the phase diagrams of the 4-junction array and and the array of alternating 3- and 6-junctions. For these  arrays the corresponding Majorana model exhibits always also longer range tunneling, which will in general be weaker in amplitude due to the $N>3$ junction geometries, which dictate that not all wire ends can be equispaced. By allowing modest control over these longer range couplings, we show that uniform $N>3$ networks can be driven into topological phase with Chern numbers $|\nu|>1$.

We again separate again the Josephson physics, encoded in the junction angle parameter $\beta$, from the tunneling couplings and define $J=S \sin \delta \phi$. In the $N>3$ junctions each wire end couples to the $N-1$ other wire ends. This means that the corresponding Majorana model will have $N-2$ different range couplings $S^{(n)}$ originating from each site. For instance, for the 4-junction array illustrated in Fig. \ref{fig:arrays}, we would denote by $S^{(1)}$ and $S^{(2)}$ the nearest  and next nearest neighbour couplings across each junction, respectively. Due to longer range couplings being in general weaker, we will consider coupling configurations where $S^{(1)} \geq S^{(2)} \geq S^{(3)} \geq \ldots$.

Figure \ref{fig:N4_beta} shows the phase diagram for the 4-junction array for $S^{(1)}=S^{(2)}$. We find it being very similar to that of the 3-junction array with only minor continuous changes as we make $S^{(2)}$ smaller than $S^{(1)}$. Thus while the corresponding square-octagon spin model is known to exhibit a rich phase diagram due to longer range interactions\cite{Kells2011}, we conclude that most of it is inaccessible by the longer range intra-junction interactions only.

\begin{figure}[ht]
\begin{tabular}{c}
\includegraphics[width=.35\textwidth,height=0.3\textwidth]{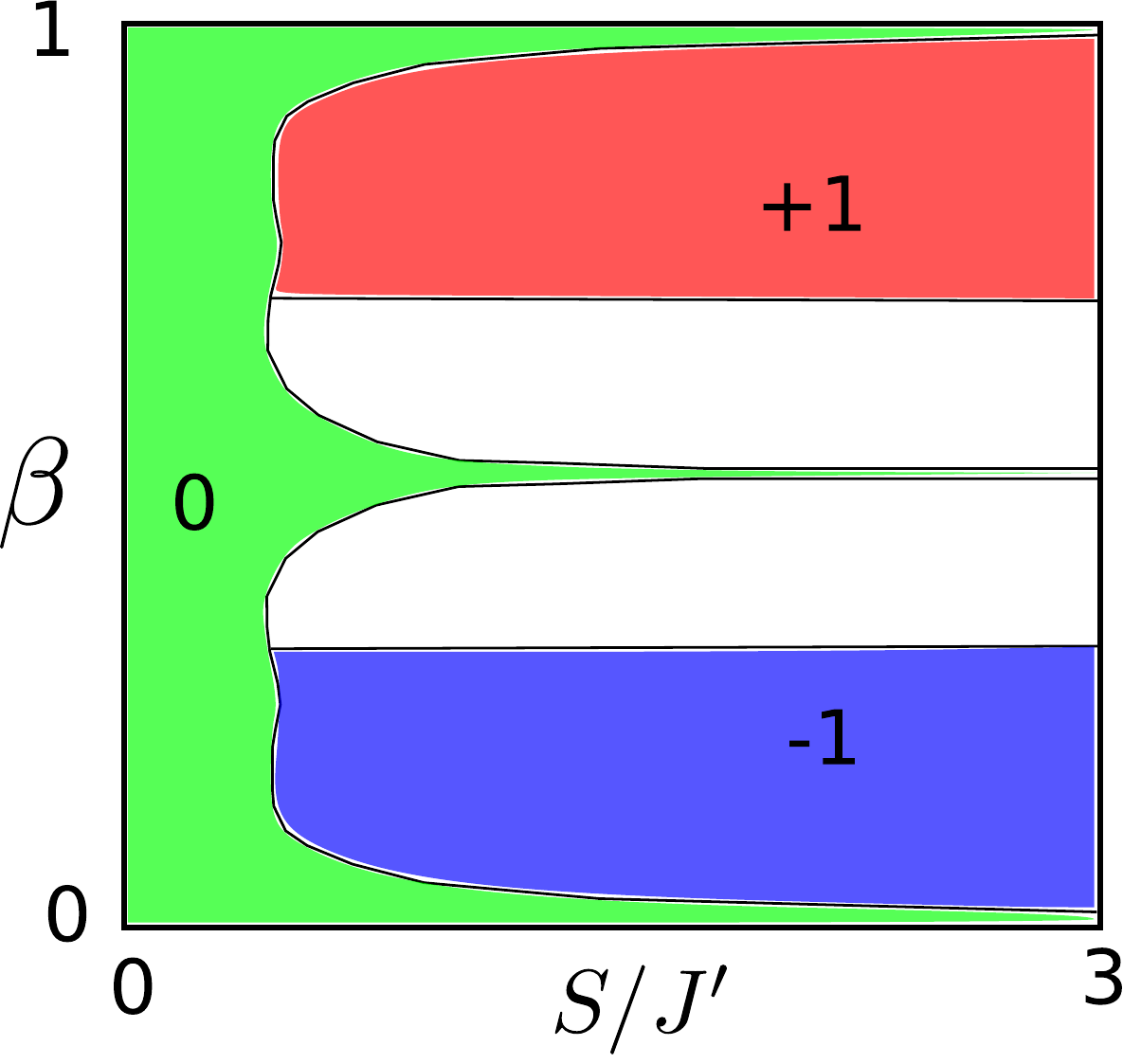} \\
(a) \\
\includegraphics[width=.35\textwidth,height=0.3\textwidth]{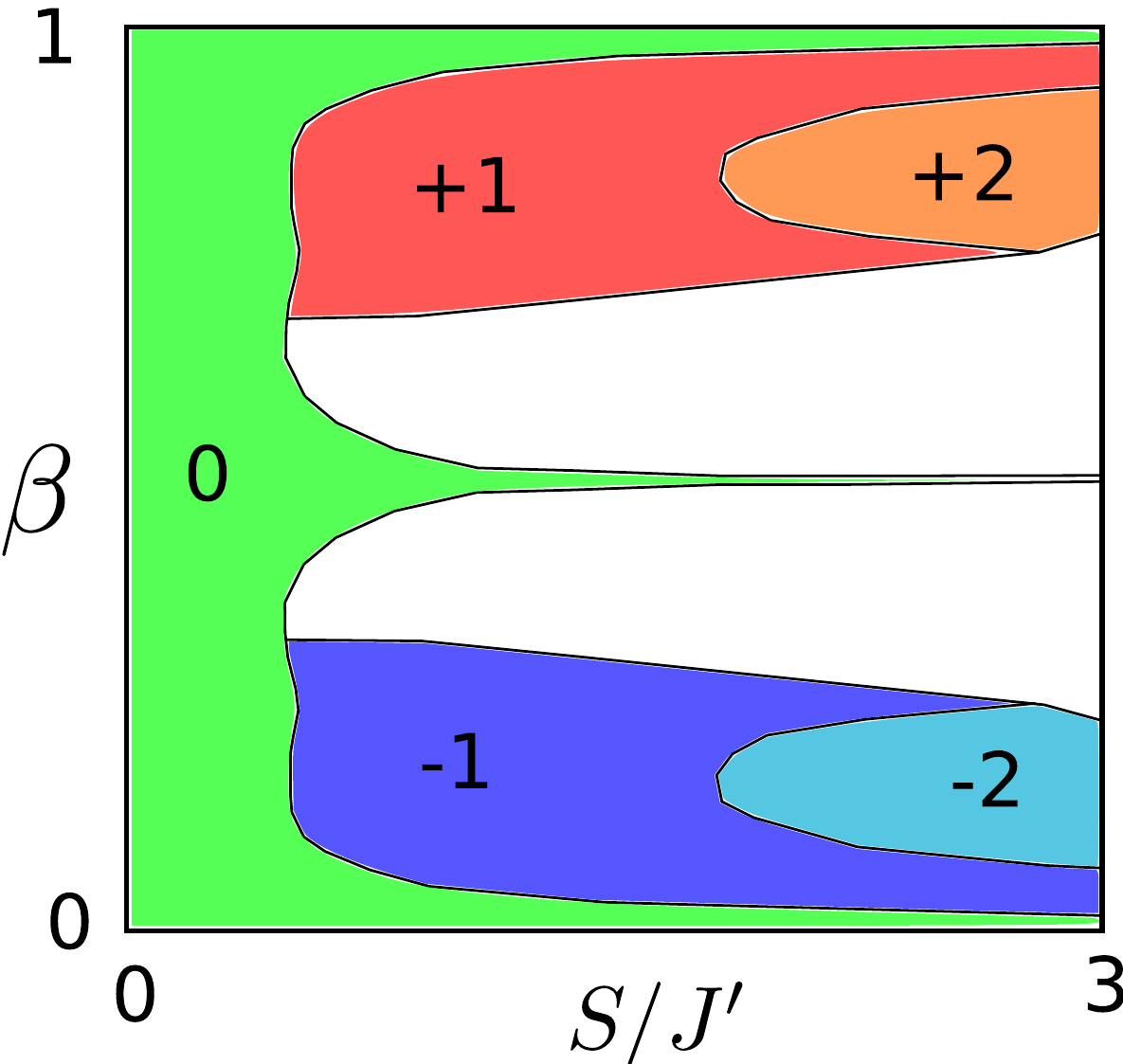} \\
(b)
\end{tabular}
\caption{The phase diagram for an array of alternating 3- and 6-junctions for (a) $S^{(1)}=S^{(2)}=S^{(3)}=S$ and (b)   $S^{(1)}=S$,$S^{(2)}=0.9S$,$S^{(3)}=0.7S$. The white regions are gapless. }
\label{fig:N6_beta} 
\end{figure}

The situation is more interesting for the array of alternating 3- and 6-junctions, as shown in Figure \ref{fig:N6_beta}. We find that when the longer range couplings decay moderately (we take here $S^{(2)} = 0.9 S^{(1)}$ and $S^{(3)} =0.7 S^{(1)}$) as predicted from increasing juntions widths, we find that collective states characterized by $\nu=\pm 2$ can emerge even in a uniform system.

\section{Fermions, Vortices and Spins}
\label{sect:FVS}

In the tri-valent Kitaev models, the spin degrees of freedom simultaneously encode both fermionic and gauge degrees of freedom. To solve the system one singles out a particular gauge/vortex sector by specifying the eigenvalues of the loop/plaquette symmetries of the model. In each sector then one finds that the remaining unknowns are described by a quadratic fermionic Hamiltonian, where the signs of the hopping amplitudes reflect the underlying vorticity and ones choice of gauge. 

The Majorana operators are a neatest way to describe these fermionic degrees of freedom. In the original solution, the Pauli algebra is formulated in terms of Majoranas in an enlarged Hilbert space \cite{Kitaev2006}. The advantage of this method is that the Majorana hopping model can be written down on the same lattice as the original spins. From here one can very quickly and accurately calculate Chern numbers and eigenspectra.  One price to pay for this simplicity is that there are apparently too many ways to create a particular vortex sector and considerable care must be taken when interpreting the eigenstates of such a system , see  Ref. \onlinecite{Pedrocchi2011}. 

\begin{figure}
\includegraphics[width=.45\textwidth,height=0.26\textwidth]{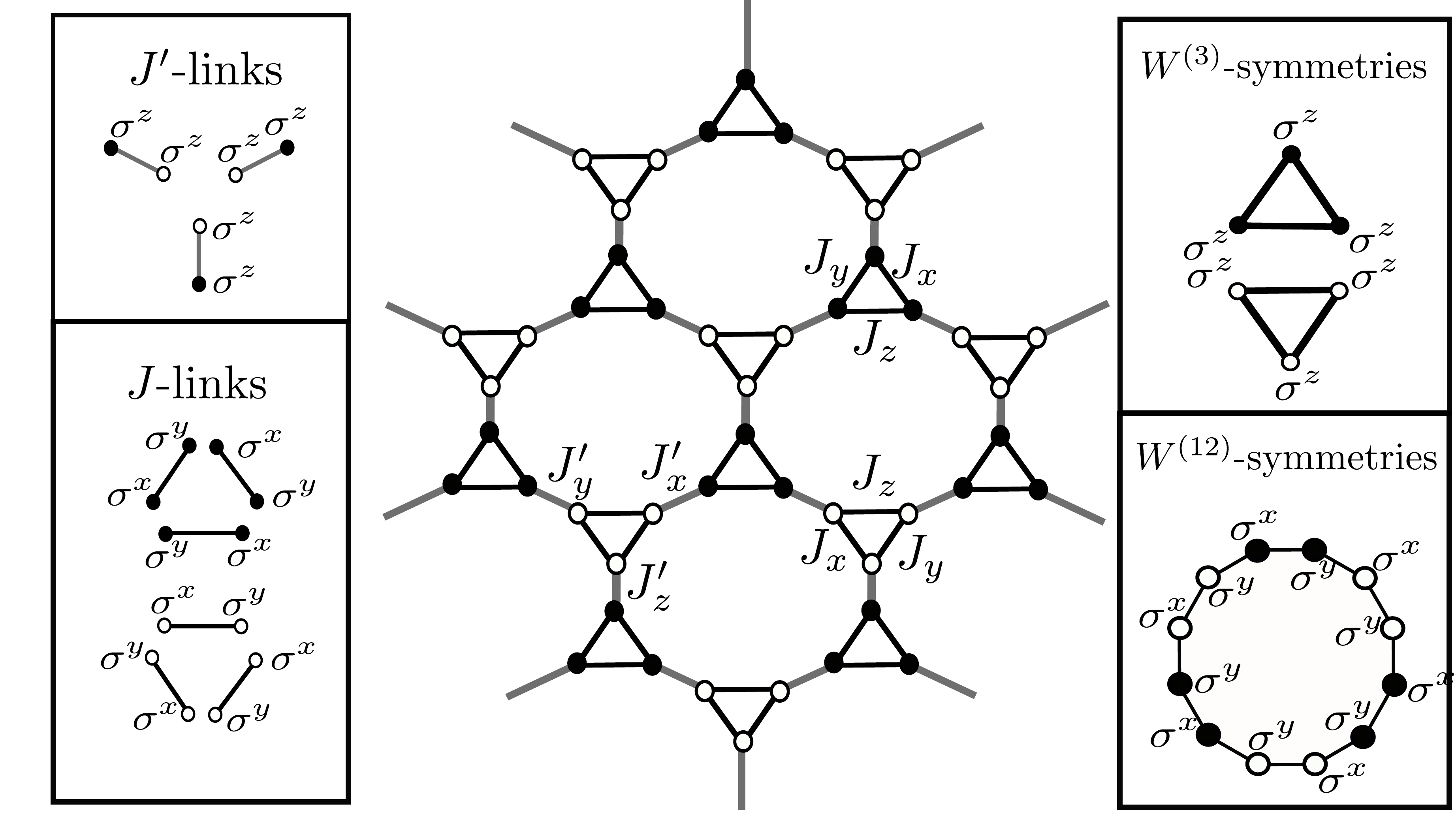}
       \caption{A basis rotation allows one to write all $J'$-links as $\sigma^z \sigma^z$ and all $J$-links as $\sigma^x \sigma^y$. The corresponding plaquette symmetries are given in the right hand panel. }
       \label{fig:basischange}
\end{figure}

Another method is to formulate the problem in terms of complex fermions \cite{Feng2007,chen2008,Kells2009,Kells2010}. In the method outlined in \cite{Kells2009,Kells2010} one first makes a local basis rotation such that all $J'$-links are of the form $\sigma^z \sigma^z$ and that all $J$-links are of the form $\sigma^x \sigma^y$ , see Figure  \ref{fig:basischange}. From here one can  identify  anti-ferromagnetic configurations of the $J'$-links with hardcore bosons and effective spin degree of freedom \cite{Schmidt2007,Dusuel2008}.  Attaching a string of spins to each hard-core boson further reduces the system to a fermionic hopping model coupled to a $\ZZ_2$ gauge field. The choice of string convention determines which gauge one uses.  Here it can also be seen that fermionic vacuum states in each sector correspond to a Toric code stabilizer states on the effective spin level \cite{Kells2009,Kells2010}.

There are not many situations where it is more advantageous to work with the spin degrees of freedom. One example however is in understanding the robustness of the system to virtual processes which at an intermediate stage involve the excitation of fermions or vortices.  This strength is exploited in the weak $J$ limit and allows the low energy sector of the full spin model to be perturbatively mapped to a toric code Hamiltonian \cite{Kitaev2001,Schmidt2007,Dusuel2008,Kells2008}.  Note that only in the 0-fermion sector can the vortex eigenvalues of the full Kitaev spin model be exactly identified with the eigenvalues of the Toric-code excitations, see for example \cite{Dusuel2008}. In the weak $J$ limit, this 0-fermionic sector actually corresponds to the ground state manifold.

\begin{figure}
\includegraphics[width=.2\textwidth,height=0.5\textwidth]{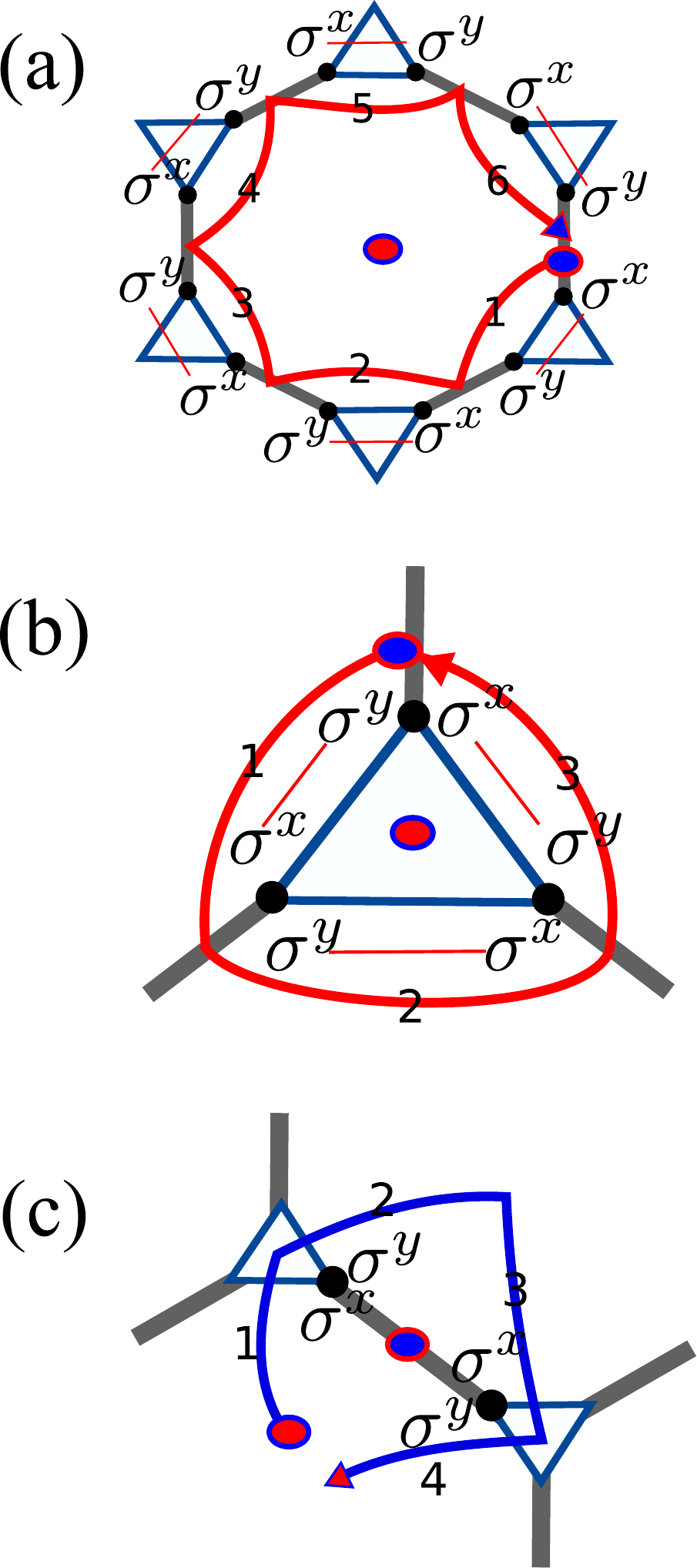}
       \caption{The spin representation offers a simple picture of both Aharonov-Bohm and Aharonov-Casher processes at the effective level. Figures (a) and (b) correspond to electron/hole motion in a closed loop. In this case we show the process for the situation where there is only one electron shared between all links. In Figure (c) we show how a vortex tunnelling through the Josephson links corresponds to the process $- \sigma^z \sigma^z = -2(c^\dagger c -I)$ measuring the parity of the enclosed wire. The branch cuts which connect vortices indicate all of the Josephson couplings connecting to the wire are changed by $-1$, corresponding with a $2 \pi $ phase slip of the superconducting phase.  }
       \label{fig:ABAC}
\end{figure}

As the Jordan-Wigner mappings allows one to, up to a sign, identify occupied local fermionic modes as anti-ferromagnetic configuration of two spins connected by a $J'$-link, this means that the creation/annihilation/motion of {\em fermions}  can be  understood as Pauli {\em bi-linear} $J$ terms $\sigma^x  \sigma^y $ terms connecting each $J'$-link to others. Thus in the spin language, closed trajectories made of the Pauli bi-linear terms that make up the Hamiltonian can always be written as a product of the plaquette symmetries and we recover the Aharonov-Bohm process where a fermion moving in a closed loop measures the flux or vorticity inside that loop, see Figure \ref{fig:ABAC} (a) and (b).  

On the other hand, because {\em single} Pauli-operators anti-commute with two of the adjoining plaquette operators, we can see that applying $\sigma^x$, $\sigma^y$, or $\sigma^x$ can be used to represent the creation/annihilation/motion of {\em vortices} in the spin model.  On the spin level then we can then also represent effective flux tunnelling through links in a loop.  In the case of a closed tunnelling path, see Figure  \ref{fig:ABAC} (c), one models the Aharonov-Casher effect where a flux loop measures the parity of the $J'$-link (wire) and changes the sign of all the Josephson tunnelling coefficients leading into the wire. The process then is equivalent to to a $2 \pi$ phase slip, see for example \cite{Pekker2013}.   

\end{document}